\documentclass[reprint,aps,prb,twocolumn,groupedaddress,nobibnotes,superscriptaddress]{revtex4-1}
\setcitestyle{numbers,square}

\usepackage[english]{babel}
\usepackage{amsmath,amssymb,amsfonts} 
\usepackage{graphicx}
\usepackage{bm}
\usepackage{xcolor}
\usepackage{color} 
\usepackage{dcolumn}
\usepackage{chngcntr}
\usepackage{hyperref}
\usepackage{mathtools}
\usepackage{amsmath}
\usepackage{amsfonts}
\usepackage{amsthm}
\usepackage{amssymb}
\usepackage{bm}
\usepackage{graphicx}
\usepackage{float}
\usepackage{multirow}
\usepackage{float}
\usepackage{placeins}
\usepackage{caption}
\usepackage{ragged2e}
\usepackage[mathlines]{lineno}
\modulolinenumbers[5]
\usepackage{commath}
\usepackage{verbatim}
\usepackage{filecontents}
\usepackage{makecell}

\begin{document}

\title{Spin fluctuations steer the electronic behavior in the FeSb$_{3}$ skutterudite}

\author{Enrico Di Lucente}
\affiliation{Theory and Simulation of Materials (THEOS) and National Centre for Computational Design and Discovery of Novel Materials (MARVEL), École Polytechnique Fédérale de Lausanne, Lausanne 1015, Switzerland}
\author{Flaviano José dos Santos}
\affiliation{PSI Center for Scientific Computing, Theory and Data, Paul Scherrer Institute, 5232 Villigen PSI, Switzerland}
\affiliation{Brazilian Center For Research in Physics, Rua Doutor Xavier Sigaud 150, Rio de Janeiro, 22290-180, Brazil}
\author{Nicola Marzari}
\affiliation{Theory and Simulation of Materials (THEOS) and National Centre for Computational Design and Discovery of Novel Materials (MARVEL), École Polytechnique Fédérale de Lausanne, Lausanne 1015, Switzerland}
\affiliation{PSI Center for Scientific Computing, Theory and Data, Paul Scherrer Institute, 5232 Villigen PSI, Switzerland}

\date{\today}

\begin{abstract}
Skutterudites are promising materials for thermoelectric and spintronics applications. Here we explore spin fluctuations in the FeSb$_{3}$ skutterudite and their effect on its electronic structure using Hubbard-corrected density-functional theory calculations. We identify multiple magnetic and charge-disproportionated configurations, with an antiferromagnetic metallic ground state. Paramagnetic fluctuations modeled through a special quasirandom spin structure open a 61 meV gap, consistent with experiments. This state features non-degenerate spin channels and band-avoided crossings, resembling a Luttinger-compensated ferrimagnet.
Mapping the electronic structure to a Heisenberg Hamiltonian fails to explain the low Néel temperature ($\lesssim$10 K), suggesting that factors such as stoichiometry and magnetic exchange frustration may play an important role, calling for more detailed experimental investigations.
\end{abstract}

\maketitle

\section{Introduction}

Research on skutterudites has focused in the last few years on novel and complex phenomena, including superconductivity \cite{shirotani1997superconductivity,b2002heavy}, long-range magnetic order \cite{danebrock1996magnetic,bauer2002crystal}, Kondo insulator behavior \cite{bauer2001kondo}, non-Fermi liquid behavior \cite{takeda2001effect,baumbach2008non}, metal-insulator transitions \cite{sekine1997metal}, and valence fluctuations \cite{dilley1998intermediate}.
Thanks to these intriguing properties the technological interest in these materials has become very high, especially for thermoelectric and spintronics applications \cite{singh1997calculated,yang2014dp,zong2017skutterudite,zhang2017realizing}. Skutterudites are known for their distinctive cage-like crystal structure, where the icosahedral cages formed by group-15 elements can be filled with rare-earth elements, lanthanides, or alkali metals \cite{sales2003filled,nolas1996effect,leithe2003ferromagnetic}. These filler atoms give rise to the so-called filled skutterudites, whose remarkable thermal properties have been extensively studied in experiments \cite{nolas1996effect,nolas1998effect,sales2000thermoelectric,sales2003filled,qiu2011},
supported by numerous first-principles calculations focusing on the electronic and vibrational properties
\cite{chaput2005transport,tang2015convergence,koza2008breakdown,di2023crossover}.
Meanwhile, transition-metal atoms outside the cages can play a critical role in enabling long-range magnetism \cite{danebrock1996magnetic, bauer2002crystal}.

Here, we focus on the FeSb$_{3}$ skutterudite, that displays a complex magnetic phenomenology.
Möchel \textit{et al.} were able to synthesize FeSb$_{3}$ in thin films \cite{mochel2011lattice} and their electrical resistivity measurements suggest a semiconducting behavior (common to other unfilled skutterudites \cite{slack1994,sofo1998electronic,kurmaev2004electronic}) characterized by a tiny electronic gap of 16.3 meV. At the same time, they infer a low-temperature paramagnetic behavior with an effective moment of 0.57$\mu$\textsubscript{B}/f.u., resulting from magnetic susceptibility measurements. This value is considerably greater than that measured for CoSb$_{3}$ ($\sim0.10\mu$\textsubscript{B}/f.u. \cite{yang2002valence}), pointing to the latter skutterudite as non-magnetic \cite{wee2010effects}. Importantly, Möchel \textit{et al.} measured the magnetic susceptibility for FeSb$_{3}$ between 10 and 300 K and found a paramagnetic Curie-Weiss behavior down to 70 K \cite{mochel2011lattice}.
Further synthesis and characterization of FeSb$_{3}$ films were performed by Daniel \textit{et al.} \cite{daniel2015structural} that derived a room-temperature conductivity about four times larger than that of Möchel \textit{et al.}, leading to a semi-metallic behavior; in this latter study, however, the magnetic properties have been explored exclusively through first-principles simulations. The differences between FeSb$_{3}$ and CoSb$_{3}$ suggest that local magnetic moments could play a major role in FeSb$_{3}$, something confirmed by the several self-consistent magnetic states we found in the present study (see later in the text).

\section{Methods}

Here, we first underline the importance for first-principles calculations to describe localized $d$-orbitals on the iron atoms accurately. In fact, the use of standard approximations in density-functional theory (DFT) \cite{DFT_1,DFT_2} tends to delocalize and over-hybridize these electrons due to self-interaction errors (SIE) \cite{kulik2006density}. This leads to especially poor predictions in mixed-valence systems, since the mixed-valence fractional elements get filled with the same fraction of electrons, thus wiping out the correct oxidation state \cite{cococcioni2019energetics,timrov2022accurate}. For this reason, we employ state-of-the-art extended Hubbard-corrections \cite{anisimov1991band,dudarev1998electron}, using orthogonalized atomic orbitals as Hubbard projectors, as implemented in the \texttt{Quantum ESPRESSO} distribution. Importantly, we determine Hubbard $U$ and $V$ parameters via linear-response \cite{cococcioni2005linear} density-functional perturbation theory (DFPT) \cite{timrov2018hubbard,timrov2021self}, ensuring a fully parameter-free and non-empirical approach. Second, we address paramagnetism with quasirandom arrangements of local magnetic moments, which allows to describe the electronic behavior of FeSb$_{3}$ at finite temperature. In particular, we use special quasirandom structures (SQS) \cite{zunger1990special} within supercell representations \cite{malyi2020false, trimarchi2018polymorphous,varignon2019origin} to provide a highly accurate description of the random paramagnetic phase \cite{wang2020understanding}. The SQS method is designed to closely replicate target pair and many-body correlation functions characteristic of disordered systems \cite{zunger1990special}, achieving this within the constraints of a given supercell size. Observables calculated from an SQS structure approximate the ensemble average for the random disordered system \cite{zunger1990special,wei1990electronic}. Furthermore, a SQS in a large supercell provides more reliable results than an ensemble average on smaller random supercells \cite{wei1990electronic,wang2020understanding}. In this context, it has been observed how a SQS-simulated cation order-disorder (alloy-like) transition can also cause fluctuations in the band gap \cite{rey2014band,gonzalez2009mechanism,veal2015band}. In this work, the SQS method was applied using the \texttt{mcsqs} code from the Alloy Theoretic Automated Toolkit (\texttt{ATAT}) \cite{van2002alloy}.

To estimate the paramagnetic transition temperature of FeSb$_3$, we map its electronic structure onto a Heisenberg Hamiltonian. 
For that, we apply the infinitesimal-rotations method based on the magnetic force theorem\,\cite{liechtenstein_local_1987}.
This approximation treats rigid infinitesimal spin rotations as perturbations using Green's function formalism. The Green's functions are determined through maximally-localized Wannier functions obtained with \texttt{Wannier90}\,\cite{quantum_espresso}, and the exchange parameter using \texttt{DFWannier}\,\cite{santos_flavianojsdfwannierjl_2024}.
We use a $25\times25\times25$ k-point mesh and 100 points in the complex energy contour to compute the exchange parameters.

The Wannierized Hamiltonian is mapped onto the following Heisenberg Hamiltonian:
\begin{equation}
    \mathcal{H} = - \frac{1}{2} \sum_{ij, \mu\nu} J_{ij}^{\mu\nu} \hat e_i^\mu \cdot \hat e_j^\nu,
\end{equation}
where $J_{ij}^{\mu\nu}$ are the magnetic exchange parameters; $i,j$ are indices for the unit cells and $\mu,\nu$ represent the various sites inside of a unit cell; and $\hat e_i^\mu$ is a unit vector along the direction of the on-site magnetic moment.
We then determine the Néel temperature, $T_N$, using the mean-field approximation (MFA), which is given by the highest eigenvalue of
\begin{equation}
    \Theta(\mathbf q) = \frac{\varepsilon}{3 k_\text{B}} J_{\mathbf q}^{\mu\nu},
\end{equation}
where $J_{\mathbf q}^{\mu\nu} = \sum_{j} J_{ij}^{\mu\nu} e^{-i \mathbf q \cdot \mathbf R_{ij}^{\mu \nu}}$\,\cite{sasioglu_exchange_2005}, $\varepsilon = 0.719$ is the reduced temperature\,\cite{garanin_self-consistent_1996}, and $\mathbf{q}$ is evaluated along a high-symmetry path of the crystal.
In addition, we perform Monte-Carlo simulations with the \texttt{Vampire package}\,\cite{evans_atomistic_2014, vampire} using a $4\times4\times4$ macrocell with periodic boundaries and an equilibration time of 2000 Monte-Carlo steps, and statistical averaging also over 2000 steps. 

\section{Results and Discussion}

According to ligand-field theory, interactions between ligand (Sb) electrons and the Fe 3$d$-orbitals split the crystal field, raising the $d_{z^{2}}$ and $d_{x^{2}-y^{2}}$ orbitals in energy, while lowering $d_{xz}$, $d_{xy}$, and $d_{yz}$, making them nondegenerate. When this energy gap is large,  Fe 3$d$-electrons occupy the lower-energy orbitals first, resulting in a low-spin (LS) state with minimal unpaired electrons. Conversely, when spin-pairing energy is higher, electrons maximize unpaired states in a high-spin (HS) configuration. In this work, we begin by determining the electronic ground state of the 16-atom primitive unit cell of FeSb$_{3}$ with \textit{Im}$\bar{3}$ space group, systematically exploring all possible self-consistent states based on various combinations of charge disproportionation and local magnetic moment orientations. Crucially, identifying the most realistic spin configuration—whether LS or HS—is vital for accurately modeling magnetism in FeSb$_{3}$. When calculations are performed without initializing the occupancy matrix of the Fe 3$d$-orbitals to a specific LS target, the system often evolves towards a HS state. In particular, the issue of HS configurations being erroneously favored in DFT+Hubbard calculations has been discussed \cite{mariano2020biased,mariano2021improved}. Specifically, in a runaway fashion, HS configurations lead to lattice expansion, which increases the linear response $U$ value, further stabilizing the HS state. In fact, the latter becomes energetically more favorable than the LS one only at large and unphysical cell expansions (see also Fig. \ref{fig:C3_eos} in Appendix \ref{EOS_appendix}). Moreover, the extremely low effective paramagnetic moment measured in FeSb$_{3}$ \cite{mochel2011lattice} suggests minimal unpaired electrons, in agreement with a LS ground state configuration. Here, all the self-consistent magnetic states obtained for FeSb$_{3}$ were generated by initializing the system with specific target LS occupations of the $d$-orbitals (see Appendix \ref{magn_constr_section}). The constraint is then removed, so that the final results correspond to self-consistent states of the unconstrained energy functional. To confirm the existence and completeness of these states, all of them have been reobtained using the \texttt{RomeoDFT} code \cite{ponet2024energy}, which performs a systematic search to explore the potential energy surface landscape shaped by the various magnetic states (including HS ones).

A summary of all the LS self-consistent magnetic states for FeSb$_{3}$ in the 16-atom primitive cell is provided in Table \ref{tab:magnetic_states} of Appendix \ref{appendix_16_atoms_cell}. Given the LS configuration, they vary according to charge disproportionation between oxidation number 2+ ($d^{6}$, non-magnetic) or 3+ ($d^{5}$, magnetic) and, in the latter case, according to the direction taken by the spin of the unpaired electron.
\begin{figure}[!htb]
\centering
\includegraphics[width=0.48\textwidth]{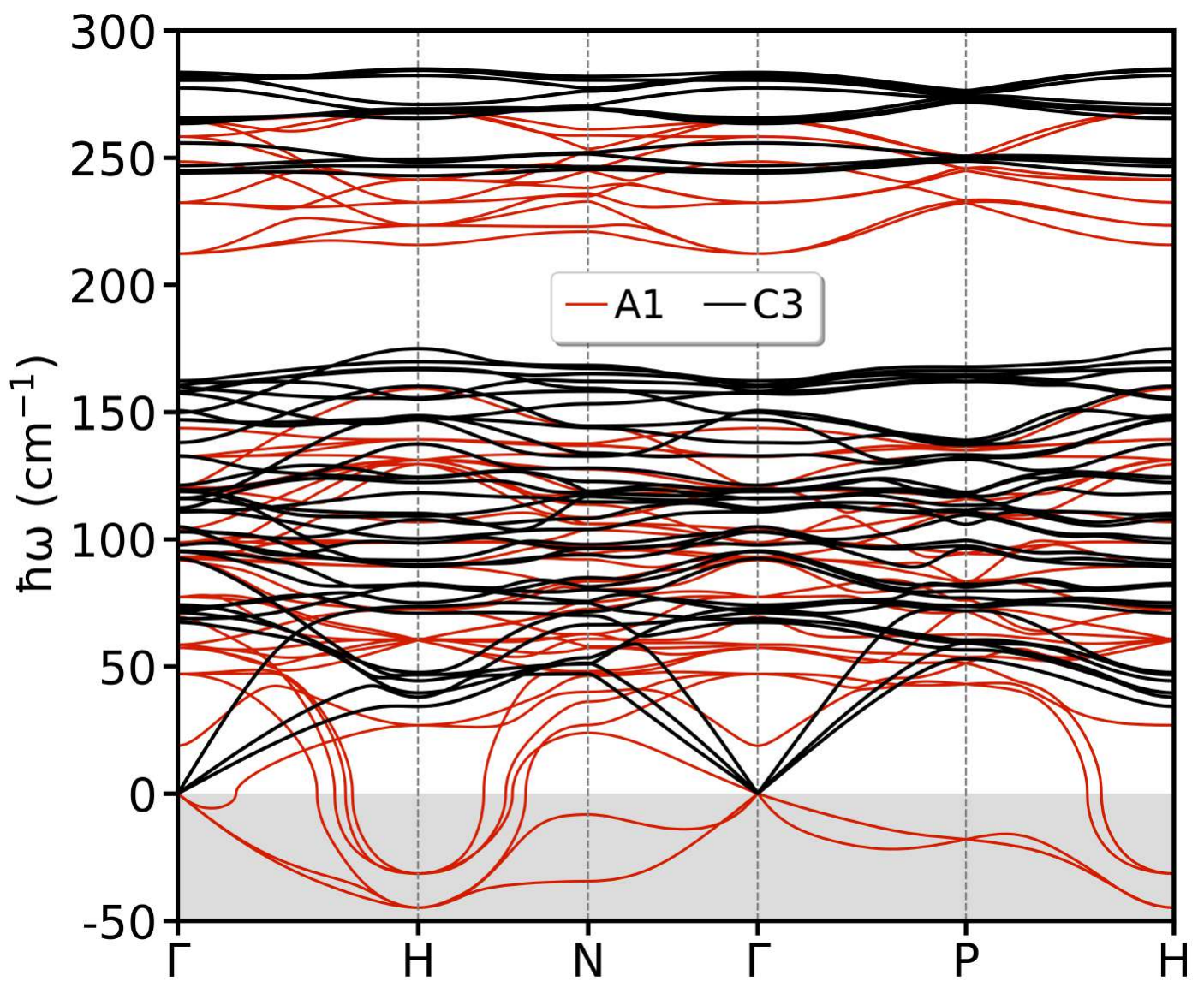}
\caption{Phonon dispersions of the non-magnetic A1 (red) solution and the antiferromagnetic C3 (black) ground state in the 16-atom primitive cell of the FeSb$_{3}$ skutterudite calculated along the main symmetry directions. The grey shaded area indicates the region of imaginary phonon frequencies.}
\label{fig:phonons}
\end{figure}
Among these states, we have one non-magnetic Fe\textsuperscript{2+}-monovalent state (A1); three ferromagnetic (B1, B2 and B3), one antiferromagnetic (B4) and one ferrimagnetic (B5) Fe\textsuperscript{2+}/Fe$^{3+}$-mixed-valence states; and one ferromagnetic (C1), one ferrimagnetic (C2) and one antiferromagnetic (C3) Fe$^{3+}$-monovalent states. The state C3 has the lowest energy and is antiferromagnetic (AFM), namely with a 1:1 ratio of Fe\textsuperscript{3+$\uparrow$} and Fe\textsuperscript{3+$\downarrow$} atoms. The other configurations with only Fe$^{3+}$ atoms are those closest in energy to the C3. Setting the energy of the ground state (C3) to 0 eV, we find the state C2 (ferrimagnetic with 3:1 ratio of Fe\textsuperscript{3+$\uparrow$} and Fe\textsuperscript{3+$\downarrow$} atoms) and the ferromagnetic state C1 at 0.32 and 0.56 eV/Fe, respectively, above the ground state. The non-magnetic A1 state has the highest energy while the mixed-valence configurations are found in an intermediate energy range. In addition, a monotonous trend with decreasing energy is obtained as the number of magnetic Fe$^{3+}$ atoms in the unit cell increases, pointing out the fundamental role played by magnetism. Moreover, given the same number of Fe$^{3+}$ atoms, antiferromagnetic states are always lower in energy than ferromagnetic ones, demonstrating the material's preference for AFM ordering (see Fig. \ref{fig:E_vs_U}). The linear-response $U$ and $V$ parameters computed for each of these states are also reported in Table \ref{tab:magnetic_states}. The value of $U$ obtained for the Fe atoms is fully consistent with those reported from linear-response calculations \cite{kulik2008self}. It is important to highlight that, in this study, the magnetic moments on Fe atoms deviate from the nominal value ($1\mu_{{\rm{B}}}$) expected for Fe in the LS state. Specifically, the DFT+Hubbard calculations indicate partial occupations of the $e_{g}$ molecular orbital, leading to a magnetic moment $>1\mu_{{\rm{B}}}$ when integrated over atomic spheres (see Fig. \ref{fig:occupationsLS_HS} in Appendix \ref{occupation_appendix}). We find that this enhancement arises from strong Fe-Sb hybridization, as evidenced by the notable linear response-computed Hubbard $V$ values, ranging from 0.35 to 0.50 eV. Previous works highlighted the relation between magnetic configuration (HS or LS) and lattice structure, showing how both can depend on the Hubbard $U$ correction \cite{hsu2009first,hsu2011spin}, for which, we reiterate, we use here and unbiased linear-response approach.

The relaxed geometries of the C3, C2, and C1 states with the 16-atom cell result in a lattice parameter equal to 9.201\AA, 9.355\AA$\,$and 9.351\AA$\,$, respectively; the first being in excellent agreement with the reference experimental measurement of 9.238\AA$\,$\cite{mochel2011lattice} and with other experiments where the value is slightly smaller (9.154\AA$\,$\cite{daniel2015structural}, and 9.176\AA$\,$\cite{hornbostel1997rational}). Fig. \ref{fig:phonons} shows the phonon dispersion relation of the C3 AFM ground state in the 16-atom primitive unit cell together with that of the A1 NM state. The C3 state clearly shows no imaginary phonon frequencies, confirming its stability, whereas the A1 state is unstable, consistent with previous findings \cite{rasander2015electronic}. We further scrutinized the stability of the different self-consistent states by computing their phonon dispersion relations. As further representative examples, Fig. \ref{fig:phonons_C1_C2_B1} in Appendix \ref{phonon_appendix} presents the phonon dispersions for the FM C1, FiM C2, and FM B1 states. Moreover, all other configurations containing at least one Fe$^{3+}$ ion exhibit only real phonon frequencies, underscoring the critical role of local magnetism in stabilizing the structure.

Since the AFM C3 state is identified as the lowest-energy magnetic configuration, we systematically explore all possible AFM phases featuring a 1:1 ratio of Fe\textsuperscript{3+$\uparrow$} and Fe\textsuperscript{3+$\downarrow$} atoms. These calculations are conducted within the 32-atom cubic conventional cell with \textit{P}23 space group, thereby broadening the search for possible spin excitations.
\begin{table}[!htp]
  \centering
  \resizebox{0.48\textwidth}{!}{\begin{tabular}{c|c|c|c|c|c}
  \hline
  \hline
     \footnotesize{State} & \makecell{\footnotesize{Valence}\\\footnotesize{Configuration}} & \footnotesize{M.O.} & \makecell{$\rm{E_{g}}$\\(meV)} & \makecell{\footnotesize{$\Delta\rm{E}$}\\\footnotesize{(eV/Fe)}} & \makecell{\footnotesize{$|\mu_{\mbox{\tiny{Fe}}}|$}\\\footnotesize{($\mu_{\mbox{\tiny{B}}}$/Fe)}} \\ 
    \hline
    $\,\,$\makecell{\footnotesize{NM}\\}& \footnotesize{8Fe\textsuperscript{2+}} & \footnotesize{NM} & metal & \footnotesize{+2.64} & \footnotesize{0.00}   \\
    \hline
    $\,\,$\makecell{\footnotesize{FM}\\}& \footnotesize{8Fe\textsuperscript{3+$\uparrow$}} & \footnotesize{FM} & metal & \footnotesize{+0.09} & \footnotesize{1.53}   \\
    \hline
    $\,\,$\makecell{\footnotesize{A}\\}& \footnotesize{4Fe\textsuperscript{3+$\uparrow$}, 4Fe\textsuperscript{3+$\downarrow$}} & \footnotesize{AFM} & metal & \footnotesize{+0.06} & \footnotesize{1.74}   \\
    \hline
    $\,\,$\makecell{\footnotesize{A\textquotesingle}\\}& \footnotesize{4Fe\textsuperscript{3+$\uparrow$}, 4Fe\textsuperscript{3+$\downarrow$}} & \footnotesize{AFM} & metal & \footnotesize{+0.06} & \footnotesize{1.74}   \\
    \hline
    $\,\,$\makecell{\footnotesize{C}\\}& \footnotesize{4Fe\textsuperscript{3+$\uparrow$}, 4Fe\textsuperscript{3+$\downarrow$}} & \footnotesize{AFM} & metal & \footnotesize{+0.04} & \footnotesize{1.85}   \\
    \hline
    $\,\,$\makecell{\footnotesize{SQS}\\}& \footnotesize{4Fe\textsuperscript{3+$\uparrow$}, 4Fe\textsuperscript{3+$\downarrow$}} & \footnotesize{“PM”} & 61 & \footnotesize{+0.03} & \footnotesize{1.75}   \\    \hline
    $\,\,$\makecell{\footnotesize{G}\\}& \footnotesize{4Fe\textsuperscript{3+$\uparrow$}, 4Fe\textsuperscript{3+$\downarrow$}} & \footnotesize{AFM} & metal & \footnotesize{0.00} & \footnotesize{1.94}   \\
  \hline
  \hline
  \end{tabular}}
  \caption{Different possible magnetic configurations in the 32-atom conventional cell with \textit{P}23 space group. We list the relevant magnetic order (M.O., FM being ferromagnetic, AFM being antiferromagnetic and “PM” being paramagnetic), electronic band gap ($\rm{E_{g}}$, when present; metal otherwise), energy difference per Fe atom with respect to G-AFM ($\Delta\rm{E}$) and average magnetic moment per Fe atom ($|\mu_{\mbox{\tiny{Fe}}}|$). Data are ordered by decreasing energy with respect to the G-AFM ground state. Here, we focus on isolating the energy differences caused by flipping magnetic moments. To achieve this, we keep the crystal structure fixed to the optimized geometry of the G-AFM ground state and apply the same $U$ and $V$ values—calculated from the linear response of the C3 state—across all configurations. The oxidation states of Fe atoms are obtained following the scheme proposed in Ref. \cite{sit2011simple}.}
  \label{tab:magnetic_states_32_atoms}
\end{table}
In this structure, Fe atoms occupy the eight vertices of a cube allowing to model conventional A-, A\textquotesingle-, C-, and G-AFM magnetic configurations \cite{kim2016revealing} (see Fig. \ref{fig:magnetic_configs_32_atoms_cell} in Appendix \ref{magnetic_configs_appendix}), distinguished by the directions of the magnetic moments of Fe atoms.
\begin{figure}[!htb]
\centering
\includegraphics[width=0.48\textwidth]{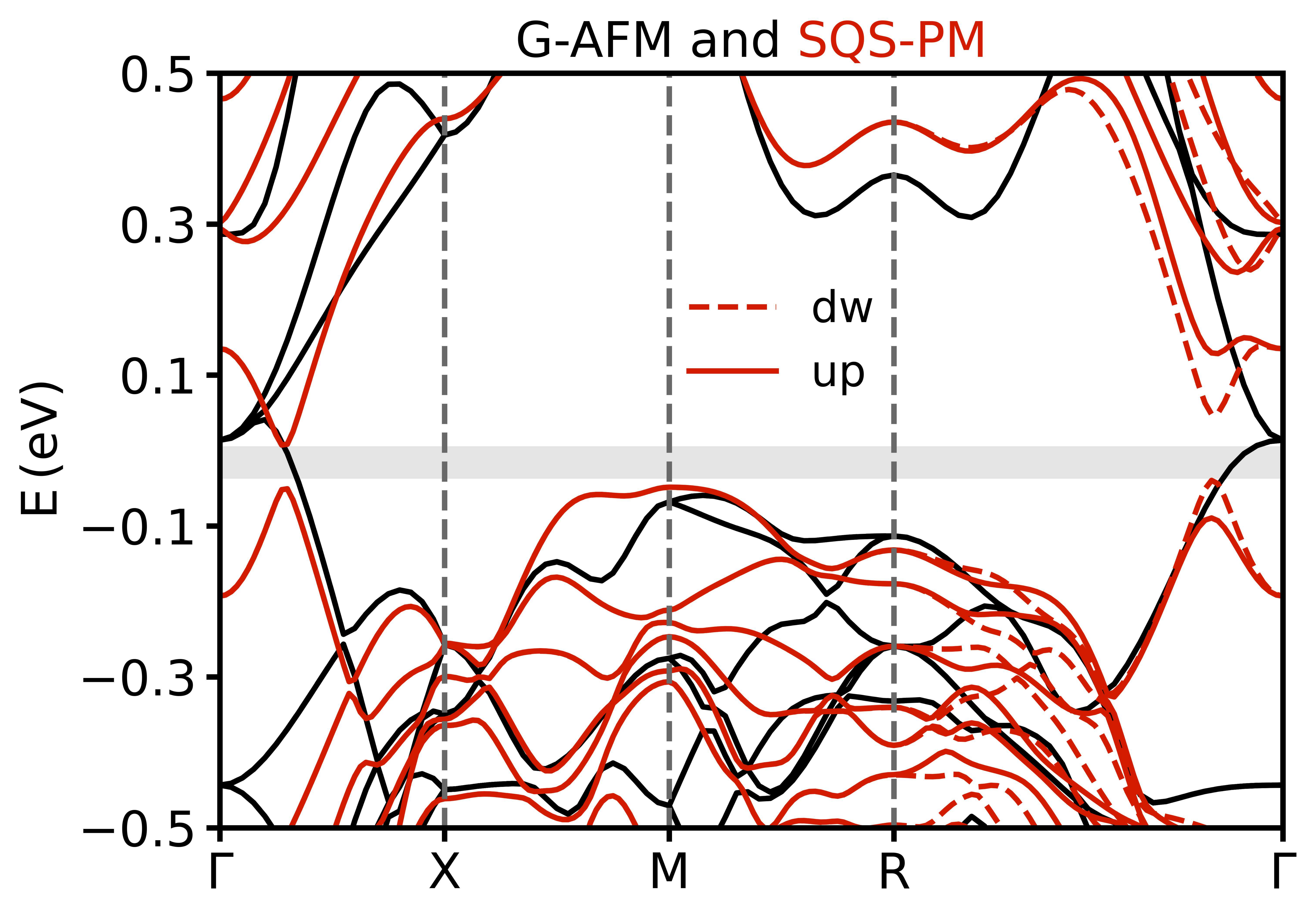}
\caption{Electronic band structure of the antiferromagnetic G-AFM (black) and “paramagnetic” SQS-PM (red) states (see Table \ref{tab:magnetic_states_32_atoms}) in the 32-atom cubic unit cell with \textit{P}23 space group. The band structures are calculated along the main symmetry directions. The spin up (spin down) contribution to the band structure is given by full (dashed) lines. The shaded gray area represents the band gap of the semiconducting SQS-PM state. The zero of the energy axis has been set equal to the Fermi level and the highest occupied level for metallic and semiconducting states respectively.}
\label{fig:bands}
\end{figure}
In addition to these four well-known AFM orderings, we construct the model for a paramagnetic state within the 32-atom cubic cell in a magnetic configuration that we denote SQS-PM (see Fig. \ref{fig:magnetic_configs_32_atoms_cell}).
Notably, the G-AFM structure is identified as the ground-state magnetic configuration, in agreement with Ref. \cite{lemal2015thermoelectric}. Table \ref{tab:magnetic_states_32_atoms} presents a comparison of the energetics, magnetic moments, and electronic behavior for all the magnetic states allowed in the 32-atom cubic cell. For completeness, the NM (dynamically unstable, see Fig. \ref{fig:phonons}) and FM phases are also calculated in the 32-atom conventional cubic cell.

In Fig. \ref{fig:bands}, we compare the electronic band structure of the SQS-PM and G-AFM states. We clearly see how the metallic behavior is replaced by a semiconducting character when we allow for spin fluctuations through the SQS approach mimicking a disordered paramagnetic arrangement of magnetic moments. In fact we observe the opening of an energy gap of about 61 meV in the SQS-PM phase, already in qualitative agreement with the experimental data obtained by Möchel \textit{et al}, given by 16.3 meV. In addition, the SQS state shows avoided crossing features similar to those of topological insulators along the $\rm{\Gamma}$-X and R-$\rm{\Gamma}$ directions.
Most interestingly, along the R-$\rm{\Gamma}$ direction, the SQS state also shows non-degenerate spin up and spin down bands, suggesting a potential transition from the G-AFM ground state to a Luttinger-compensated \cite{smolyanyuk2024tool1,mazin2022altermagnetism} ferrimagnetic (FiM) state. Luttinger-compensated ferrimagnets represent a fascinating and often overlooked class, sometimes mistakenly linked to the concept of altermagnetism \cite{mazin2022altermagnetism,vsmejkal2022emerging,yuan2020giant,yuan2021prediction,vsmejkal2022giant,gonzalez2023spontaneous,gonzalez2025spin} or spin-flipping antiferromagnets \cite{ma2021multifunctional}. In contrast to altermagnets, which rely on crystal symmetry for magnetic compensation, Luttinger-compensated ferrimagnets are not symmetry-compensated \cite{mazin2022altermagnetism}. They also differ from spin-density waves \cite{gruner1994dynamics}, which are not symmetry-compensated but exhibit magnetic structures incommensurate with the lattice. Instead, Luttinger-compensated ferrimagnets possess magnetic structures that are commensurate with the lattice and can be divided into spin-up and possibly spin-down sublattices, with the total magnetization being either zero (or “fully compensated”) or a nonzero integer value, according to Luttinger's theorem \cite{luttinger1960fermi,luttinger1960ground}. When the total magnetization is zero, it is strictly zero and remains unaffected by small perturbations such as pressure or strain \cite{mazin2022altermagnetism}. This is the case for the SQS state. 
The possibility of a Luttinger-compensated ferrimagnetic phase is particularly intriguing, as such magnetic phase could help explain the peculiar experimental results on magnetic susceptibility \cite{mochel2011lattice}. Finally, it is worth noting that this analysis was performed using the conventional 32-atom cell. However, if we consider the 16-atom primitive cell of FeSb$_{3}$, it would accommodate altermagnetic states \cite{smolyanyuk2024tool1}.

The electronic band structure and density of state of all the 32-atom magnetic states are given in Fig. \ref{fig:AFM_configs_bands}. As can be seen, the material's electronic behavior is substantially similar across the different phases, with the notable exception of a significant change in the bands in the SQS-PM state around the Fermi level. Moreover, we observe that the G-AFM ground state exhibits metallic behavior. In contrast, B1WC hybrid functional calculations from Ref. \cite{lemal2015thermoelectric} indicate that, although the G-AFM state remains the lowest energy configuration, it behaves as an insulator with a band gap of around 33 meV. This result is in good agreement with the band gap observed in the SQS phase in the present work. While in Ref. \cite{lemal2015thermoelectric} the insulating behavior seems to naturally emerge alongside antiferromagnetism, the underlying physics in the present case is quite different. Here, instead of studying the system as an ordered antiferromagnet, we allow for a quasirandom network of local magnetic moments using an SQS approach to capture the paramagnetic nature of the material, which is expected to persist even at very low temperatures \cite{mochel2011lattice}. It is also worth noting that in Ref. \cite{lemal2015thermoelectric} even the ordered FM phase is semiconducting, and that the lattice parameter of the G-AFM phase obtained with B1WC hybrid functional greatly underestimates the reported experimental values.

To explore the transition from the G-AFM ground state to the SQS-PM state, we investigated the underlying spin-flipping mechanism involving the unpaired electron on the Fe$^{3+}$ atoms. In doing so, we identified two additional self-consistent states with intermediate energies between the two configurations (see Fig. \ref{fig:very_first_excitations} in Appendix \ref{magnetic_configs_appendix}). The first, just 74 meV higher in energy with respect to the ground state, can be described as a G-AFM structure exhibiting what we refer to as “internal delocalization”. According to the eigenvalues of the occupation matrix within the Hubbard manifold, the unpaired electron of one Fe$^{3+}$ atom is equally delocalized between two atomic orbitals that form part of the $t_{2g}$ manifold (see Fig. \ref{fig:very_first_excitations}).

The second intermediate state between the G-AFM and SQS-PM configurations is what we refer to as the “flip G-FiM” state. This ferrimagnetic state, located 199 meV above the ground state in energy, features a nonzero total net magnetization. It is characterized by a complete spin flip of the Fe atom that previously exhibited internal delocalization in the first intermediate state (see Fig. \ref{fig:very_first_excitations}).
Finally, flipping the magnetic moment of just one additional Fe atom, the system transitions from the flip G-FiM state to the SQS-PM state. All intermediate states between the G-AFM ground state and the SQS-PM state exhibit metallic behavior. 
Regarding the state showing “internal delocalization”, we clearly see the limitations of a single-determinant approach such as DFT. A multi-determinant approach, such as methods like complete active space self-consistent field (CASSCF) \cite{roos1980complete,siegbahn1981complete}, configuration interaction (CI) \cite{roothaan1951new}, multireference perturbation theory \cite{andersson1992second}, density matrix renormalization group (DMRG) \cite{wouters2014density}, or multireference coupled cluster (MR-CC) \cite{bartlett2007coupled}, would be necessary to capture static correlations.

Last, we construct a SQS in a 128-atom 2$\times$2$\times$2 supercell, starting from the 16-atom primitive cell with \textit{Im}$\bar{3}$ space group, which improves randomness with respect to the SQS in the 32-atom cell, thus better describing spin fluctuations of the paramagnetic state. Going from the C3 to the SQS in the 32-atom cell and finally to the SQS in the 128-atom supercell, it is clear how the approximation of the paramagnetic phase becomes regularly refined: the system has magnetic moments located in an antiferromagnetic arrangement in the ground state; when the temperature is increased the system becomes a disordered paramagnet \cite{mochel2011lattice} in which the same magnetic moments are present but follow a disordered distribution. The electronic band structure of the SQS-128 phase (see Fig. \ref{fig:sqs-128} in Appendix) shows a reduced band gap of about 38 meV, even closer to the experimental value.

\begin{figure}[!t]
\centering
\includegraphics[width=0.49\textwidth]{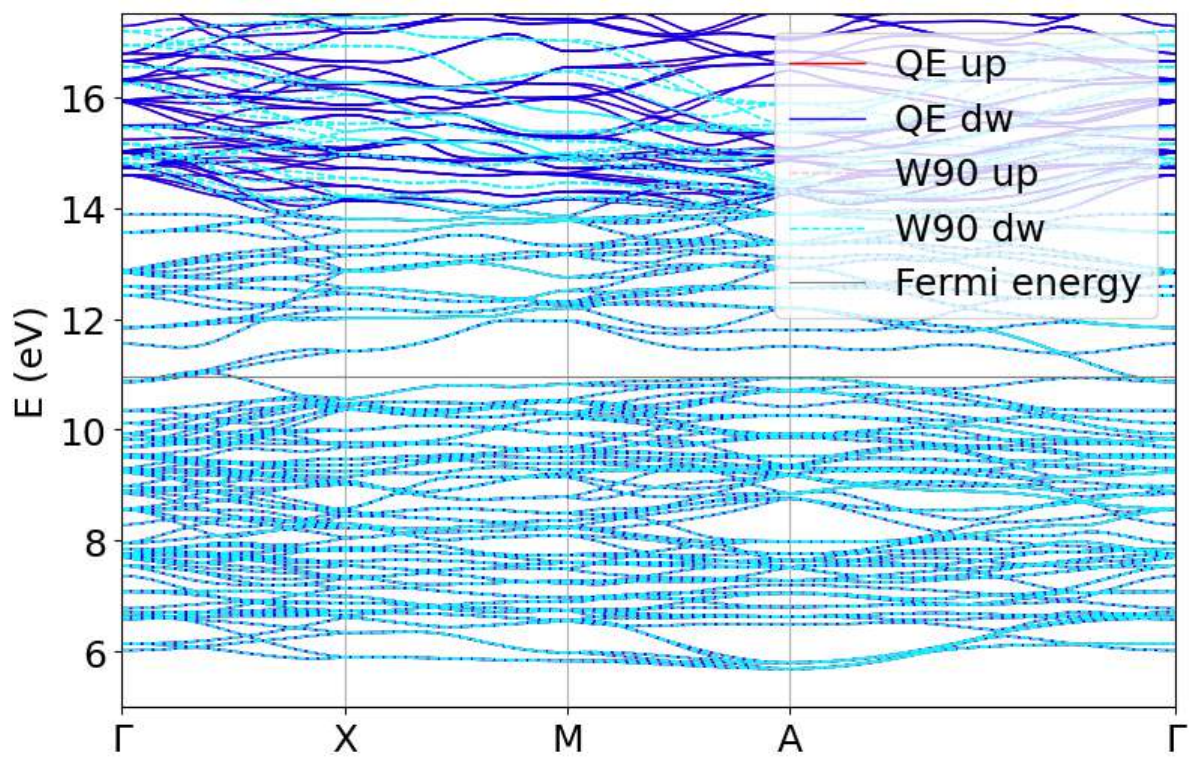}
\caption{Band structure of the G-AFM ground state of FeSb$_3$. The solid lines correspond to the DFT results, and the dashed line to the Wannier interpolated bands. 
\label{fig:BANDS}}
\end{figure}

Experimentally, the magnetic properties of FeSb$_3$ have been studied by Möchel \textit{et al.}\,\cite{mochel2011lattice}. In their work, as previously mentioned, a paramagnetic behavior of the magnetic susceptibility was observed down to 10\,K. Fitting the susceptibility with the Curie-Weiss law (between 70 and 300\,K), they obtained an effective magnetic moment of 0.57(6) $\mu_\text{B}$ and a Curie-Weiss temperature of $\sim0$\,K.
It is noteworthy that the effective moment of CoSb$_3$, a compound belonging to the same family as FeSb$_3$, is 0.10(6)\,$\mu_\text{B}$\,\cite{friak2018origin}.
Doping this material with Fe, $\text{Co}_{1-x}\text{Fe}_x\text{Sb}_3$, with $x$ from 0 to 0.1, yields an increasing effective moment of up to 1.7\,$\mu_\text{B}$.
Obviously, at the limit $x=1$, FeSb$_3$, the effective moment is much smaller: 0.57\,$\mu_\text B$.
Results by Yang \textit{et al.}\,\cite{friak2018origin} on Fe$_2$AlTi further illustrate how sensitive a material's magnetism can be to its stoichiometry.
In our DFT (PBEsol+U+V) results with Hubbard parameters determined from first principles, we obtained a magnetic moment integrated on a sphere around the Fe atoms of 1.94 $\mu_\text{B}$ in the G-AFM ground state and about 1.79 $\mu_\text{B}$ in the SQS-PM state. Such high magnetic moments are consistent with the first-principles calculations by Lemal \textit{et al.}\,\cite{lemal2015thermoelectric} using hybrids exchange-correlation functions.
In the study where the experimental magnetic moment of FeSb$_3$ is reported, Möchel \textit{et al.}\,\cite{mochel2011lattice} report a stoichiometry of FeSb2.88(5) from x-ray diffraction measurements, which could, among other things, explain the discrepancy between the experimental data and the results of the first-principle calculations.

\begin{figure}[!t]
\centering
\includegraphics[width=0.45\textwidth]{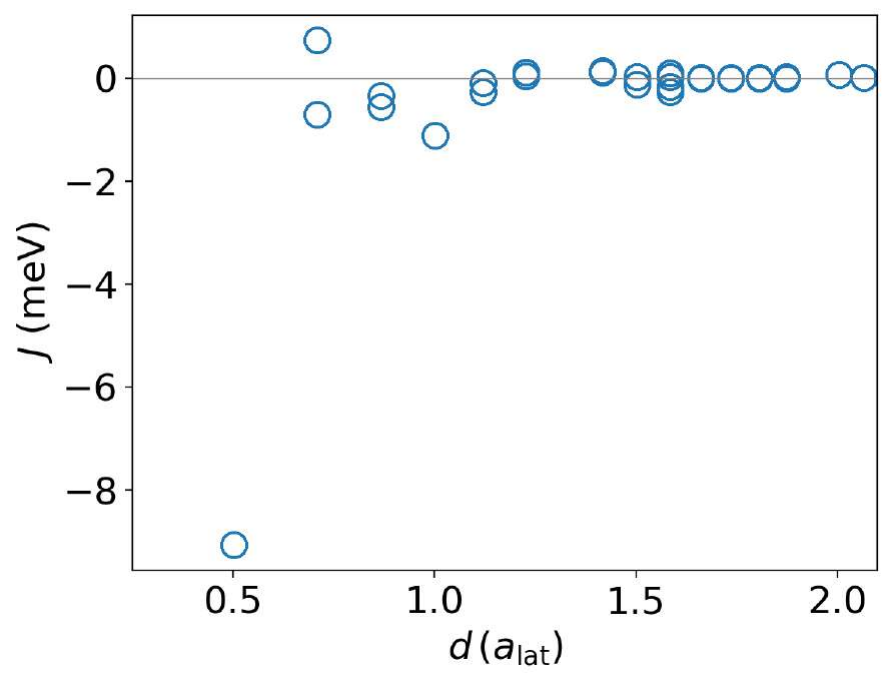}
\caption{Magnetic exchange coupling $J$ as a function of distance between coupled spins $d$. Negative values, which characterize most couplings, correspond to antiferromagnetic interactions.
\label{fig:EXCHANGE}
}
\end{figure}

\begin{figure}[!th]
    \includegraphics[width=0.47\textwidth]{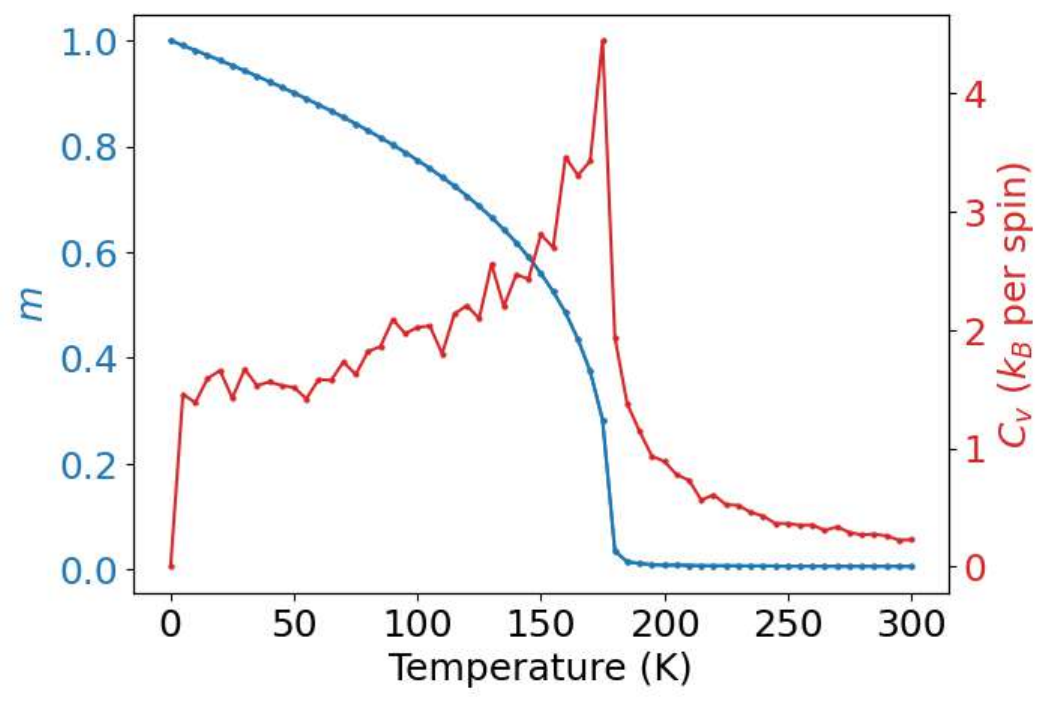}
\caption{Normalized magnetization $m$ (in blue) and the specific heat $C_v$ (in red) as a function of temperature. Their cusps are indicative of a phase transition into the paramagnetic phase with increasing temperatures.
\label{fig:HEAT}
}
\end{figure}

To estimate the paramagnetic transition temperature of FeSb$_3$, we compute the magnetic exchange interactions to parametrize a Heisenberg Hamiltonian. 
We focus on the G-AFM ground state in the 32-atom unit cell.
As the initial projections for the Wannierization, we chose the $p$ orbitals for Sb and $s$, $p$, $d$ for Fe, which allowed for a good Wannierization also of the conductance band states near the Fermi energy, as seen in Fig.\,\ref{fig:BANDS}.
The resulting exchange parameters are summarized in Fig.\,\ref{fig:EXCHANGE}, where we notice that the nearest-neighbor interactions are predominantly antiferromagnetic.
The different Sb environments make Fe pairs at the same interatomic distance inequivalent, which causes a splitting of the exchange interactions.
From this set of interactions, we obtain a corrected mean-field Néel temperature of $T_N^{\text{MFA}} = 163$\,K.
From our Monte-Carlo simulations, we obtain $T_N^{\text{MC}} = 178$\,K, which roughly matches the mean-field estimate.
Fig. \ref{fig:HEAT} shows the resulting magnetization per atom as a function of the temperature and the system's specific heat. Fitting the Monte-Carlo magnetization to the Curie-Weiss law, we obtain an effective moment of 0.8\,$\mu_\text{B}$.
Although this result is closer to the experimental measurement value (0.57\,$\mu_\text{B}$), the Néel temperature is still greatly overestimated.

This system's small critical temperature $\lesssim$ 10\,K could be explained either by exchange couplings that are much smaller than those predicted by our calculations, frustration, a mismatch in stoichiometry as discussed above, or a combination of these factors. Firstly, we perform further DFT calculations using the Korringa-Kohn-Rostoker method\,\cite{rusmann_judftteamjukkr_2022} to compute the exchange parameters from the infinitesimal-rotations approach. Here, we employ the SLDA (no Hubbard U) with and without spin-orbit coupling to the G-AFM configuration.

We also obtain a high magnetic moment of about 1.6\,$\mu_\text{B}$ and a Monte-Carlo critical temperature above 120 K with an effective moment of 1.1\,$\mu_\text{B}$. Finally, we also map the total energy difference between the ground-state antiferromagnetic configuration and a ferromagnetic metastable state onto a nearest-neighbor Heisenberg model.
We obtain $T_N = 19$ K, similar to a total energy difference calculation with QE@PBEsol without U or V (14 K).
When including U+V from linear response, our calculations indicate exchange parameters at least one order of magnitude higher than that required for a Néel temperature $\lesssim$ 10\,K.

\begin{figure}[!b]
    \includegraphics[width=0.48\textwidth]{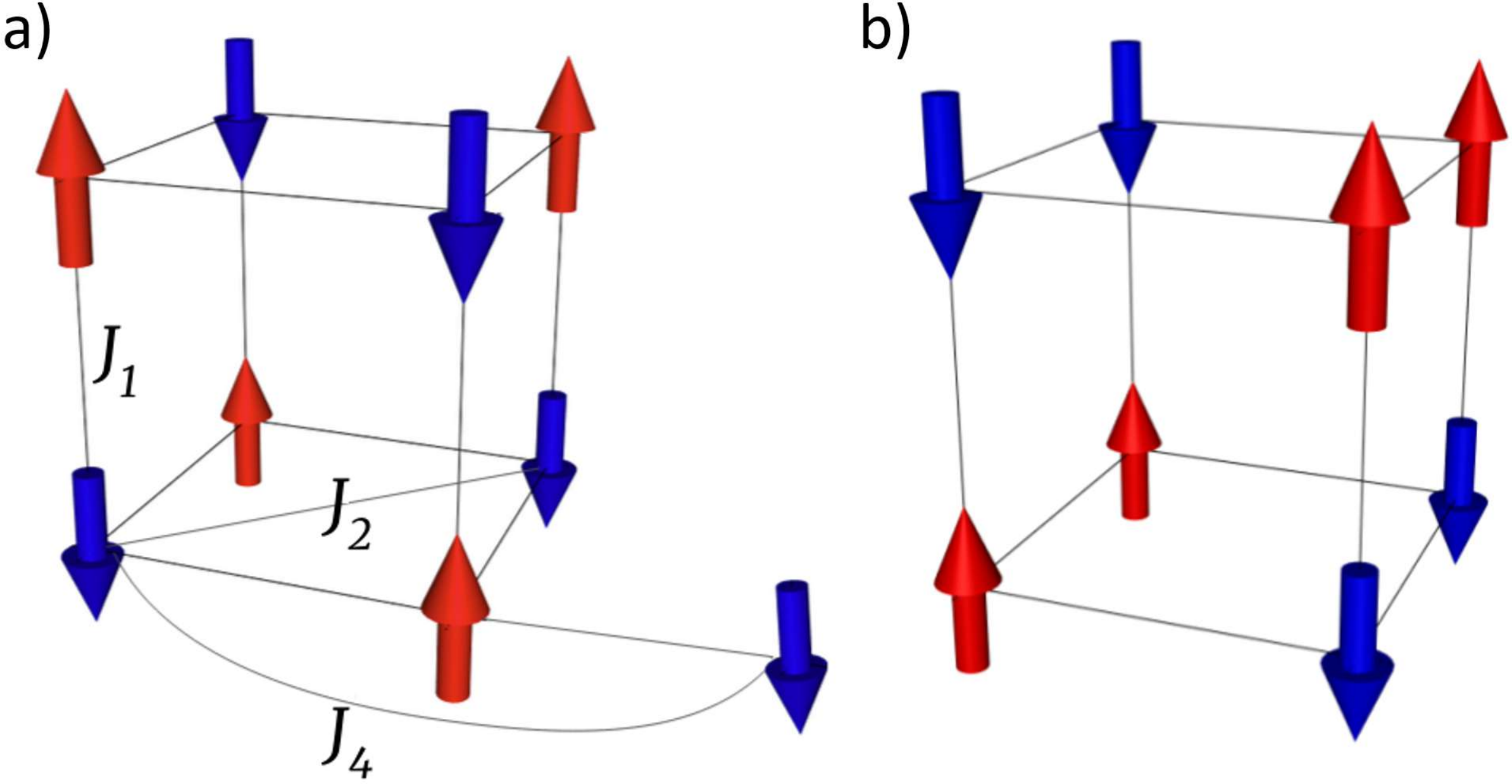}
    \caption{(a) The G-AFM spin configuration and the nearest-neighbor couplings of the cubic structure formed by the iron atoms of FeSb$_3$.
    (b) The D-AFM spin configuration induced by a dominant antiferromagnetic $J_2$ coupling.
    \label{fig:frustration_model}}
\end{figure}

In the scenario where exchange interactions are large, as suggested by our DFT@PBEsol+U+V, frustration could also lead to a relatively small $T_{N}$.
Consider a model with only the nearest-neighbor coupling $J_1=-14$\,meV.
We obtain the G-AFM ground state shown in Fig.\,\ref{fig:frustration_model}\,(a) and $T_{N}=230$\,K from Monte Carlo simulations (similar to the mean-field prediction of 234 K after corrections).
If we include an antiferromagnetic $J_2$, this will create geometric frustration because no magnetic configuration can simultaneously satisfy the $J_1$ and $J_2$ couplings. The maximum frustration is achieved at about $J_2 \sim J_1 /4$. Larger values of $J_2$ will induce a new ground state, D-AFM (named so because all parallel magnetic moments are connected by a cube's diagonal) shown in Fig.\,\ref{fig:frustration_model}\,(b). Near this point of frustration, neither the G-AFM nor the D-AFM is very dominant and the system can easily fluctuate between these two states.
Hence, $T_N$ gets much reduced from 230\,K to $\sim50$\,K for $J_2 = -3.50$\,meV (here, the mean-field theory does very poorly: $T_N = 117$\,K).
Further frustration can be induced by $J_4$.
With $J_4=-2$\,meV, long-range order is almost fully destroyed and $T_N$ reduces further.
Our DFT calculations predict that the first 5 nearest-neighbor couplings are antiferromagnetic ($J_2$ is both ferro and antiferromagnetic depending on the relative alignment of the spins), which indicates that frustration can play an important role for FeSb$_3$. Nevertheless, we do not obtain the precise ratio between, e.g., $J_1$ and $J_2$ that maximizes the frustration in this material. This can be related to the accuracy of our methods or, again, to the possible difference in stoichiometry between simulations and experiments. Future experimental works addressing whether this material will ever be ordered at temperatures below 10\,K and the sensitivity of its magnetism to the stoichiometry will be crucial to advancing our understanding.

Concerning the role of off-stoichiometric concentrations, we provide some basic assessment. Using the 32-atoms conventional unit cell containing eight Fe atoms, we compare the total energies of the G-AFM ground state and the FM metastable state. We then map the resulting energy difference onto a nearest-neighbor Heisenberg model. For the pristine FeSb$_{3}$ system, we find an exchange interaction $J=-15.39$ meV and a mean-field Néel temperature $T_{N}=257$ K.
To study the effect of stoichiometry, we remove one Sb atom from the 24 present in the conventional unit cell, yielding FeSb$_{2.875}$, which is very close to the experimentally observed FeSb$_{2.88}$ \cite{mochel2011lattice}. Due to periodic-boundary conditions, the resulting structure remains crystalline rather than forming a random alloy. To account for different local environments, we perform three independent calculations, each with a different Sb site removed. The calculated $T_{N}$ values are 80, 82, and 86 K, approximately one-third of the pristine system’s critical temperature. While this model involves strong simplifications—particularly the nearest-neighbor-only assumption, despite our other calculations suggesting a more complex set of interactions—we believe these results illustrate the significant impact of stoichiometry, demonstrating that deviations from ideal composition can substantially lower the critical temperature in this case.

\section{Conclusions}

In conclusion, the present study highlights the importance of Hubbard-corrected DFT to explore the full range of self-consistent magnetic states in FeSb$_{3}$. We also show that using the SQS approach to model the paramagnetic distribution of magnetic moments can capture the electronic behavior observed experimentally. This methodology could be also applied to other potential paramagnetic semiconductors where a higher-symmetry unit cell is inadequate for revealing hidden features, including Luttinger-compensated ferrimagnetic behavior.
Moreover, the occurrence of LS “internal delocalization” of $t_{2g}$ electrons on Fe atoms emphasizes the need of multi-determinant methods to accurately capture static correlations. Our findings identify FeSb$_{3}$ as a promising candidate for harnessing both charge and spin degrees of freedom in the development of novel electronic devices for spintronics, opening up new avenues for technological applications beyond thermoelectric energy conversion, common to many skutterudites.

Furthermore, we calculate the magnetic exchange interactions, predicting a Néel temperature of 163 K, significantly higher than the experimental observations. 
This discrepancy could be attributed to possible differences in stoichiometry between theoretical and experimental systems, although magnetic frustration could play a crucial role in explaining the lower observed paramagnetic critical temperature. Moreover, the internal delocalization of an unpaired electron within the $t_{2g}$ molecular orbital further suggests that magnetic excitations in FeSb$_{3}$ may not be those of a Heisenberg system, and might need multi-reference methods beyond single determinant approaches to accurately account for static correlations. Future experimental and theoretical research could further investigate these aspects and refine the understanding of the this material's fascinating magnetic properties.

\section{Acknowledgments}

E.D. acknowledges the Swiss National Science
Foundation (SNSF) through Grant No. CRSII5\_189924 (“Hydronics” project). N.M. acknowledges NCCR MARVEL, a National Centre of Competence in Research, funded by the Swiss National Science Foundation (Grant No. 205602)

%\bibliographystyle{apsrev4-1}
%\bibliography{bib}
%\end{document}

\appendix

\section{16-atom primitive cell with bcc (\textit{Im}$\bar{3}$) space group} \label{appendix_16_atoms_cell}

\begin{figure}[!htb]
\centering
\includegraphics[width=0.49\textwidth]{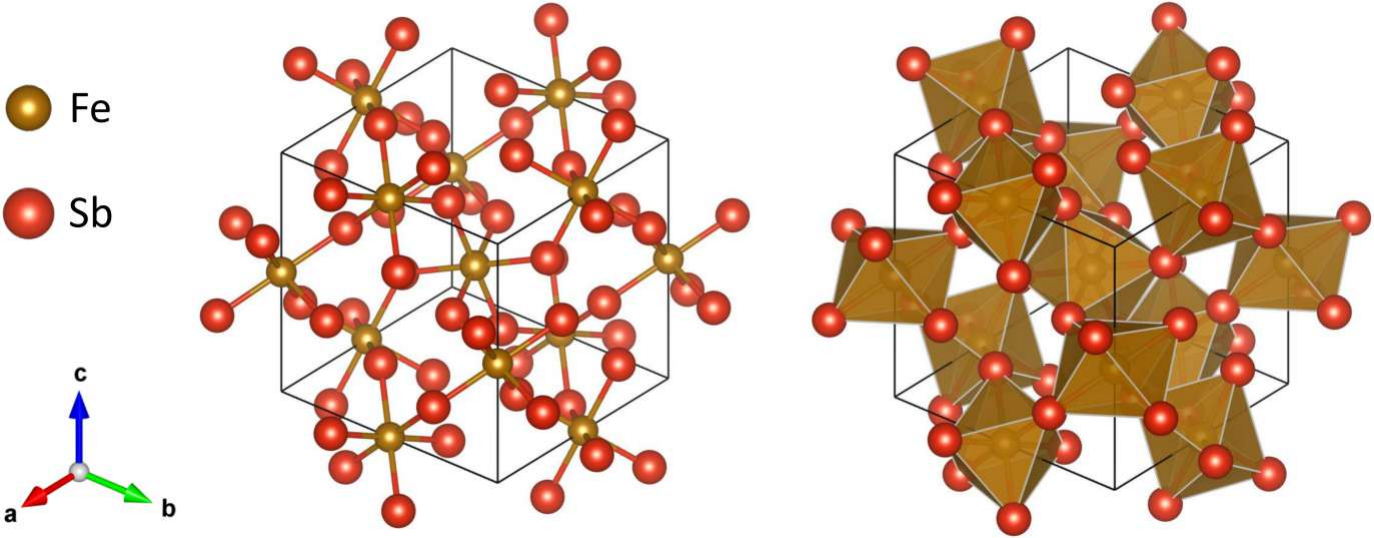}
\caption{Primitive 16-atom (\textit{Im}$\bar{3}$) unit cell of FeSb$_{3}$. Gold (red) spheres represent Fe (Sb) atoms. Interatomic Fe-Sb bonds and Fe-centered octahedral complexes are highlighted.}
\label{fig:16_atoms_cell}
\end{figure}

The 16-atom primitive cell of FeSb$_{3}$ with bcc (\textit{Im}$\bar{3}$) space group is depicted in Fig. \ref{fig:16_atoms_cell}

\section{32-atom conventional cell with simple cubic (\textit{P}23) space group}

\begin{figure}[!htb]
\centering
\includegraphics[width=0.49\textwidth]{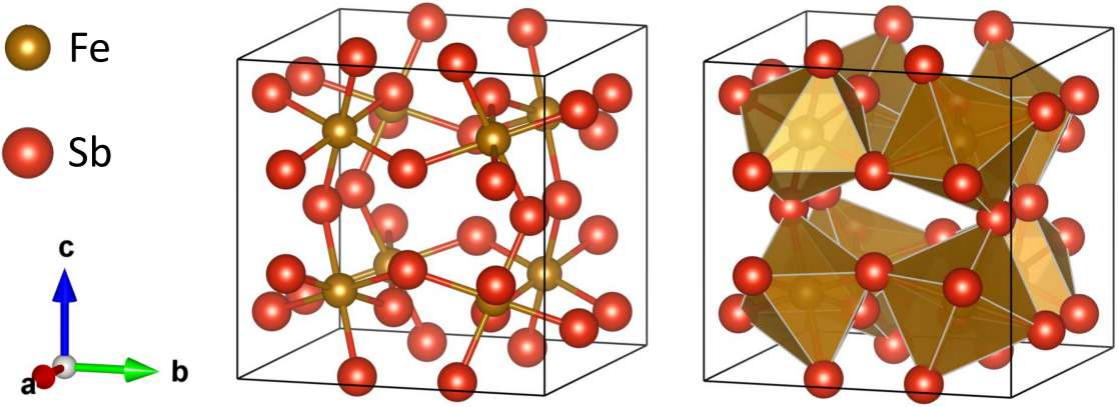}
\caption{Conventional 32-atom (\textit{P}23) cell of FeSb$_{3}$. Gold (red) spheres represent Fe (Sb) atoms. Interatomic Fe-Sb bonds and Fe-centered octahedral complexes are highlighted.}
\label{fig:32_atoms_cell}
\end{figure}
The 32-atom conventional cell of FeSb$_{3}$ with simple cubic (\textit{P}23) space group is depicted in Fig. \ref{fig:32_atoms_cell}.

\subsection{DFT+Hubbard functional and low spin initialization} \label{magn_constr_section}

When exploring the energy landscape of different magnetis states \cite{ponet2024energy}, in order to prevent the system from falling into unphysical “runaway” HS configuration, we initialize out calculation in a LS state as described below. In the DFT+U+V energy functional,
\begin{table*}[!htp]
  \centering
  \resizebox{0.75\textwidth}{!}{\begin{tabular}{c|c|c|c|c|c|c|c}
  \hline
  \hline
     \footnotesize{State} & \makecell{\footnotesize{Valence}\\\footnotesize{Configuration}} & \footnotesize{M.O.} & \makecell{$\rm{E_{g}}$\\(meV)} & \makecell{\footnotesize{$\Delta\rm{E}$}\\\footnotesize{(eV/Fe)}} & \makecell{\footnotesize{$|\mu_{\mbox{\tiny{Fe}}}|$}\\\footnotesize{($\mu_{\mbox{\tiny{B}}}$/Fe)}} & \makecell{\footnotesize{U$_{\mbox{\tiny{sc}}}$, V$_{\mbox{\tiny{sc}}}$}\\\footnotesize{(eV)}} & $a$ (\AA) \\
    \hline
    $\,\,$\footnotesize{A1} & \footnotesize{4Fe\textsuperscript{2+}} & \footnotesize{NM} & metal & \footnotesize{+2.72} & \footnotesize{0.00} & \footnotesize{6.01, 0.45} & 9.148 \\
    \hline
    \rule{0pt}{4ex} 
    \footnotesize{B1} & \makecell{\footnotesize{1Fe\textsuperscript{3+$\uparrow$}}\\\footnotesize{3Fe\textsuperscript{2+}}} & \footnotesize{FM} & metal & \footnotesize{+1.46} & \footnotesize{0.36} & \makecell{\footnotesize{4.83, 0.44}\\\footnotesize{5.64, 0.47}} & 9.295 \\
    \rule{0pt}{4ex} 
    \footnotesize{B2} & \makecell{\footnotesize{2Fe\textsuperscript{3+$\uparrow$}}\\\footnotesize{2Fe\textsuperscript{2+}}} & \footnotesize{FM} & metal & \footnotesize{+1.18} & \footnotesize{0.75} & \makecell{\footnotesize{4.70, 0.43}\\\footnotesize{5.46, 0.50}} & 9.314 \\
    \rule{0pt}{4ex} 
    \footnotesize{B3} & \makecell{\footnotesize{3Fe\textsuperscript{3+$\uparrow$}}\\\footnotesize{1Fe\textsuperscript{2+}}} & \footnotesize{FM} & metal & \footnotesize{+1.09} & \footnotesize{0.95} &  \makecell{\footnotesize{4.74, 0.43}\\\footnotesize{5.52, 0.45}} & 9.322 \\ 
    \rule{0pt}{4ex}
    \footnotesize{B4} & \makecell{\footnotesize{1Fe\textsuperscript{3+$\uparrow$}, 1Fe\textsuperscript{3+$\downarrow$}}\\\footnotesize{2Fe\textsuperscript{2+}}} & \footnotesize{AFM} & metal & \footnotesize{+0.92} & \footnotesize{0.65} & \makecell{\footnotesize{4.70, 0.44}\\\footnotesize{5.46, 0.46}} & 9.216 \\
    \rule{0pt}{4ex} 
    \footnotesize{B5} & \makecell{\footnotesize{2Fe\textsuperscript{3+$\uparrow$}, 1Fe\textsuperscript{3+$\downarrow$}}\\\footnotesize{1Fe\textsuperscript{2+}}} & \footnotesize{FiM} & metal & \footnotesize{+0.78} & \footnotesize{0.92} & \makecell{\footnotesize{4.68, 0.44}\\\footnotesize{5.33, 0.47}} & 9.273 \\
    \hline
    $\,\,$\footnotesize{C1} & \footnotesize{4Fe\textsuperscript{3+$\uparrow$}} & \footnotesize{FM} & metal & \footnotesize{+0.56} & \footnotesize{1.74} & \footnotesize{4.76, 0.36} & 9.351 \\
    $\,\,$\footnotesize{C2} & \footnotesize{3Fe\textsuperscript{3+$\uparrow$}, 1Fe\textsuperscript{3+$\downarrow$}} & \footnotesize{FiM} & metal & \footnotesize{+0.32} & \footnotesize{1.88} & \footnotesize{4.65, 0.37} & 9.355 \\
    $\,\,$\footnotesize{C3} & \footnotesize{2Fe\textsuperscript{3+$\uparrow$}, 2Fe\textsuperscript{3+$\downarrow$}} & \footnotesize{AFM} & metal & \footnotesize{0.00} & \footnotesize{1.94} & \footnotesize{4.25, 0.44} & 9.201 \\
    \hline
    $\,\,$\makecell{\footnotesize{SQS-128}\\\footnotesize{(C3)}} & \footnotesize{16Fe\textsuperscript{3+$\uparrow$}, 16Fe\textsuperscript{3+$\downarrow$}} & \footnotesize{“PM”} & 38 & \footnotesize{-0.02} & \footnotesize{1.79} & \footnotesize{4.25, 0.44} &  9.201 \\
  \hline
  \hline
  \end{tabular}}
  \caption{Self-consistent magnetic states for FeSb$_{3}$ (Fe\textsubscript{4}Sb\textsubscript{12} is the primitive unit cell), ordered by decreasing energy, with C3 being the ground state. We list the relevant magnetic order (M.O., FM being ferromagnetic, AFM being antiferromagnetic and FiM being ferrimagnetic), electronic band gap ($\rm{E_{g}}$, when present; metal otherwise), energy difference per Fe atom with respect to C3 ($\Delta\rm{E}$), average magnetic moment per Fe atom ($|\mu_{\mbox{\tiny{Fe}}}|$), self-consistent Hubbard U and V parameters derived from the calculations (U$_{\mbox{\tiny{sc}}}$, V$_{\mbox{\tiny{sc}}}$) and computed lattice parameter (to be compared with the experimental value given by 9.238 \AA$\,$\cite{mochel2011lattice}). The different magnetic states are grouped into four classes: the first row has the state with Fe\textsuperscript{2+} atoms, the second part has the mixed-valence states (both Fe\textsuperscript{2+} and Fe$^{3+}$ atoms) and the third part has states with only Fe$^{3+}$ atoms. The last part has the SQS structure obtained starting from the C3 geometry: a 2$\times$2$\times$2 supercell (of the primitive cell) with 128 atoms.}
  \label{tab:magnetic_states}
\end{table*}
\begin{equation} \label{DFT+U+V_functional}
\begin{split}
E_{\mbox{\scriptsize{DFT+U+V}}}[n(\boldsymbol{r})]=&\,E_{\mbox{\scriptsize{DFT}}}[n(\boldsymbol{r})]+\\
&+\frac{1}{2}\sum_{I\sigma}U^{I}\mbox{Tr}[(\boldsymbol{1}-\boldsymbol{\mbox{n}}^{II\sigma})\boldsymbol{\mbox{n}}^{II\sigma}]-\\
&-\frac{1}{2}\sum_{IJ\sigma}V^{IJ}\mbox{Tr}[\boldsymbol{\mbox{n}}^{IJ\sigma}\boldsymbol{\mbox{n}}^{JI\sigma}],
\end{split}
\end{equation}
the explicit expression of the occupation matrix $\boldsymbol{\mbox{n}}$ is given as follows:
\begin{equation} 
n^{I\sigma}_{mm'}=\sum_{\nu,\boldsymbol{k}}f_{\nu \boldsymbol{k}}^{\sigma}\langle\psi^{\sigma}_{\nu\boldsymbol{k}}\phi_{m'}^{I}\rangle\langle\phi_{m}^{I}\psi^{\sigma}_{\nu\boldsymbol{k}}\rangle,
\end{equation}
where $\nu$ is the band index, $\boldsymbol{k}$ is the wave vector within the Brillouin Zone, $f_{\nu \boldsymbol{k}}^{\sigma}$ are the occupations of Kohn and Sham levels, $\psi^{\sigma}_{\nu\boldsymbol{k}}$ are the Kohn and Sham wavefunctions and $\phi_{m'}^{I}$ are localized orbitals (orthogonalized atomic orbitals \cite{timrov2021self}, in the present case). Intuitively the total occupation of localized $d$ states at site $I$ is
\begin{equation} \label{occ_matr}
    n^{I}=\sum_{m\sigma}n^{I\sigma}_{mm}.
\end{equation}
In the case of Fe $d$ orbitals there will be 5 eigenvalues of the occupation matrix \eqref{occ_matr} (5 as the number of $d$ orbitals) for each spin channel. Having one eigenvalue close to 1 means that the related $d$ orbital is occupied by one electron with a given spin. On the contrary, eigenvalues different and far from 1 symbolize that the orbital is empty. As mentioned in the main article, we followed the scheme proposed in Ref. \cite{sit2011simple} to discriminate between different magnetic states, that is, to understand whether a $d$ orbital is filled or not.

Practically, we initialize the self-consistent DFT+U+V calculation with an occupation matrix having eigenvalues defined by the target LS state. After the first DFT iteration, the constraint is released and the system evolves towards the desired LS magnetic states (see Table \ref{tab:magnetic_states}), so avoiding the unphysical HS configurations.
\begin{figure}[!htb]
\centering
\includegraphics[width=0.49\textwidth]{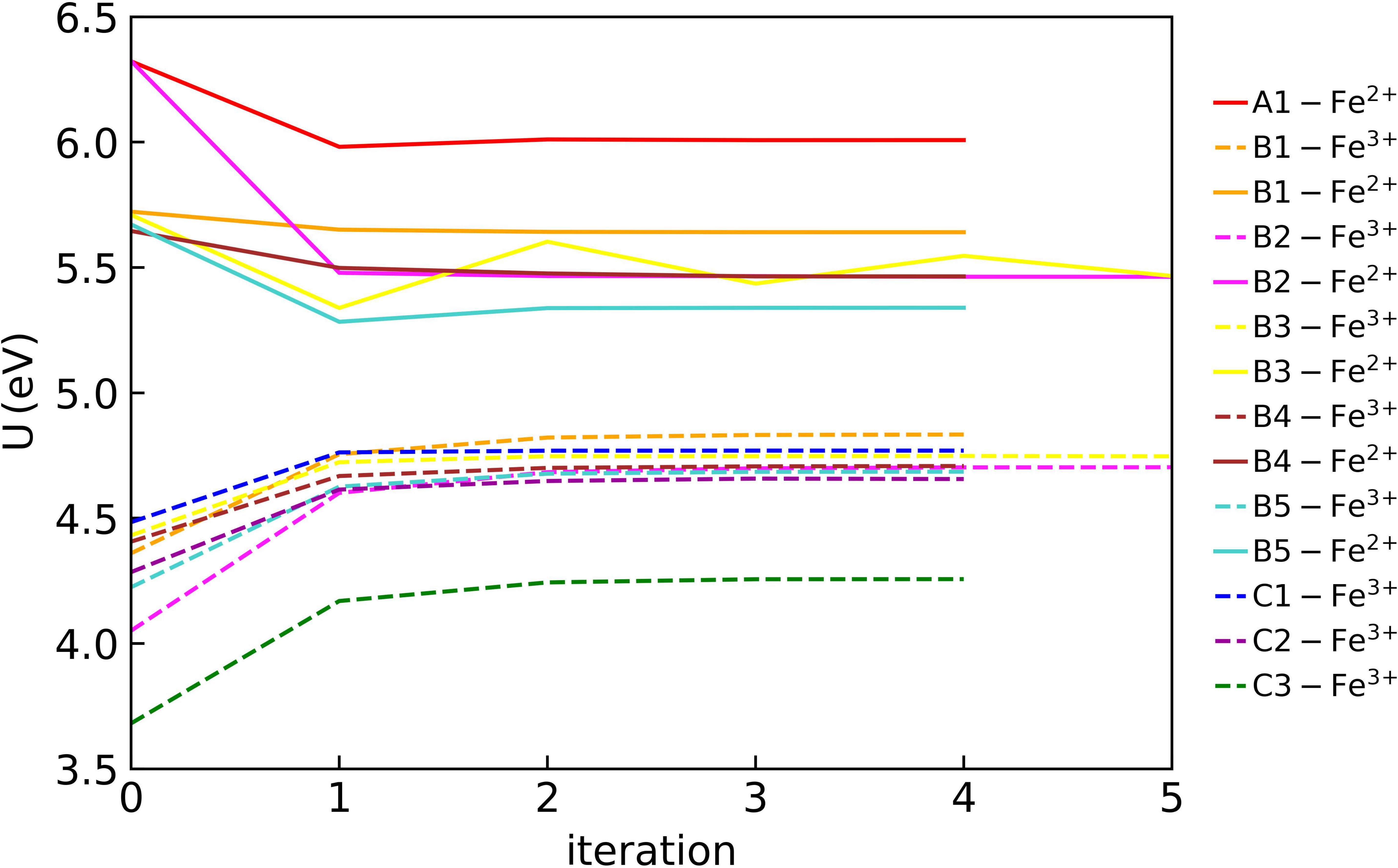}
\caption{Linear response cycle performed to obtain the $U$ values for the states discussed in Table \ref{tab:magnetic_states}. The $V$ parameters, also reported in Table \ref{tab:magnetic_states}, are not shown here in order to keep the figure uncluttered.}
\label{fig:U_values}
\end{figure}
This is equivalent to instructing the code to start the DFT self-consistency cycle from a point already close to what we believe is the local minimum of the magnetic energy landscape—namely, the LS state—while avoiding convergence toward the local minimum associated with the HS state.
\begin{figure*}[!htb]
\centering
\includegraphics[width=0.99\textwidth]{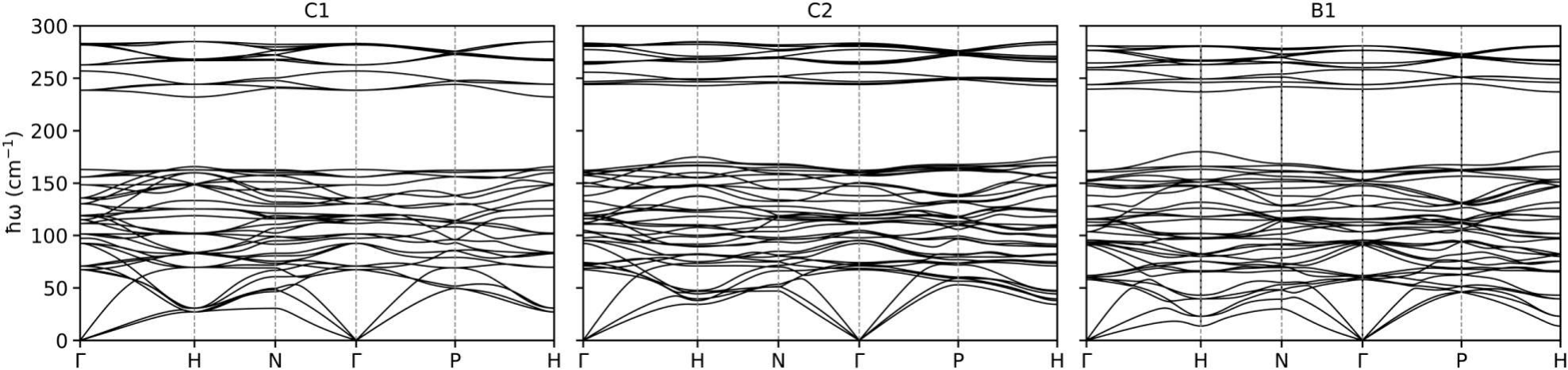}
\caption{From left to right: phonon dispersions of the ferromagnetic C1, antiferromagnetic C2 and ferromagnetic B1 states of the FeSb$_{3}$ skutterudite calculated along the main symmetry directions.}
\label{fig:phonons_C1_C2_B1}
\end{figure*}
\begin{figure*}[!htb]
\centering
\includegraphics[width=0.75\textwidth]{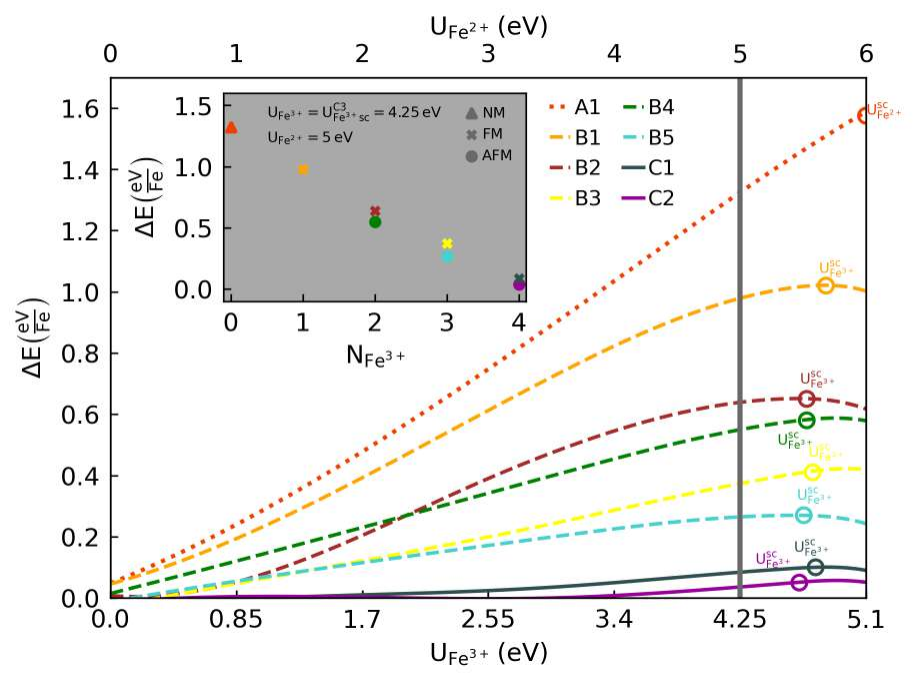}
\caption{Total energy difference per Fe atom ($\rm{\Delta E}$) between the different magnetic states listed in Table \ref{tab:magnetic_states} and the C3 state versus given values of Hubbard U of Fe\textsuperscript{2+} (upper $x$ axis) and Fe$^{3+}$ (lower $x$ axis) atoms. The ratio $\rm{U_{Fe^{2+}\,sc}}$ = $1.18\rm{U_{Fe^{3+}\,sc}}$ (to the second decimal place) applies to all the mixed-valence states. We use a dotted line for the A1 non magnetic state, dashed lines for the mixed-valence Bi states (with i=1,..,5) and full lines for the magnetic Ci states (with i=1,2,3) which have only Fe$^{3+}$ atoms. The self-consistent $\rm{U^{sc}_{Fe^{3+}}}$ values are given by circles of the same color as the curve of the related magnetic state; for the A1 state we also show the $\rm{U^{sc}_{Fe^{2+}}}$ value. The gray line represents $\rm{U_{Fe^{3+}}}$ = $\rm{U^{C3}_{Fe^{3+}\,sc}}$ = $4.25\,\rm{eV}$. The inset shows the energy difference along the gray line versus the number of Fe$^{3+}$ atoms in the primitive cell; the colors of the symbols are consistent with those of the main figure.}
\label{fig:E_vs_U}
\end{figure*}
Finally, it is very important to strengthen the fact that all the final state are self-consistent states of the unconstrained energy functional. This approach highlights the importance of identifying the most physically realistic spin configuration (either LS or HS) in DFT+Hubbard calculations. As noted in the main text, without such guidance, the system tends to relax into HS local minima, which are energetically higher than the LS configurations across the range of magnetic states studied. For completeness, in Fig. \ref{fig:U_values} we provide the linear response $U$ values for the states shown in Table \ref{tab:magnetic_states}.

\subsection{Phonon dispersions of the self-consistent magnetic states} \label{phonon_appendix}

In this section we show the phonon dispersions of further prototype self-consistent magnetic states of FeSb$_{3}$ in the 16-atom primitive cell with \textit{Im}$\bar{3}$ space group (see Table \ref{tab:magnetic_states}), namely the C1, C2 and B1 states (see Fig. \ref{fig:phonons_C1_C2_B1}). Note that the phonon dispersions of the A1 and C3 states are already given in the main article. A significant difference exists between the phonon dispersion characteristics of the C1 and C2 states compared to the B1 state. Specifically, in the B1 state, the presence of non-magnetic Fe\textsuperscript{2+} ions at a 3:1 ratio with magnetic Fe$^{3+}$ ions leads to an enhanced phonon sound velocity within the 150–200 cm\textsuperscript{-1} frequency range. This adjustment aligns the phonon dispersion of the B1 state more closely with that observed in the diamagnetic compound CoSb$_{3}$ \cite{guo2015thermal,wee2010effects}. Furthermore, the phonon dispersion of the B1 state reveals that the presence of even a minimal amount of magnetic Fe$^{3+}$ atoms is essential to avoid imaginary phonon frequencies.

\subsection{Energetics of the self-consistent magnetic states at different $U$ values}

In Fig. \ref{fig:E_vs_U} we present the energy difference per Fe atom between the states listed in Table \ref{tab:magnetic_states} and the C3 state, plotted against various values of $\rm{U_{Fe^{3+}}}$ and $\rm{U_{Fe^{2+}}}$. It can be observed that the C3 state is the one with the lowest energy for all values of $\rm{U_{Fe^{3+}}}$ ranging from 0 to 5.1 eV. Very interestingly, from the results of the mixed-valence states, we see
that $\frac{\rm{U_{Fe^{2+}\,sc}}}{\rm{U_{Fe^{3+}\,sc}}}\simeq 1.16$-$1.18$ ratio (to the second decimal place) for all these states. Therefore it is interesting to see how the energetics of all the states evolves when keeping $\frac{\rm{U_{Fe^{2+}\,sc}}}{\rm{U_{Fe^{3+}\,sc}}}=1.18$. It is evident from the inset of Fig. \ref{fig:E_vs_U} that, as the number of Fe$^{3+}$ atoms increases, the energy decreases, when U approaches its self consistent value. Moreover, at each given set of $\rm{U_{Fe^{3+}}}$ and $\rm{U_{Fe^{2+}}}$, the AFM states are always lower in energy compared to the FM ones.

\subsection{Low spin and high spin EOS of C3 state} \label{EOS_appendix}

\begin{figure}[!htb]
\centering
\includegraphics[width=0.49\textwidth]{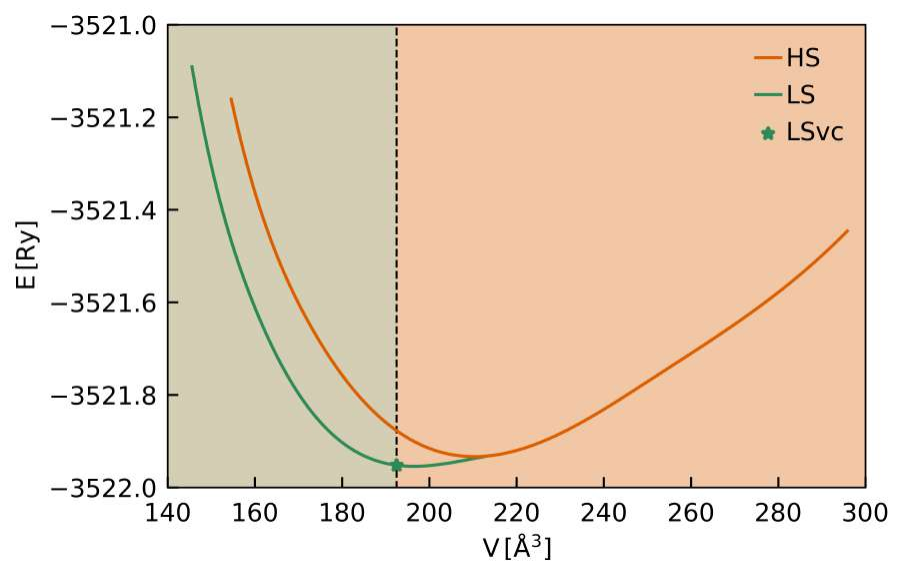}
\caption{Equation of state (energy vs unit cell volume) of the C3 LS and HS states in the the 16-atom \textit{Im}$\bar{3}$ primitive cell. The green star indicates the LS global minimum with its related optimized volume.}
\label{fig:C3_eos}
\end{figure}

In Fig. \ref{fig:C3_eos} we show the equation of state (EOS) of the C3 LS and HS states in the 16-atom \textit{Im}$\bar{3}$ primitive cell. Note that here we have used the linear response Hubbard U value of the C3 LS state (see Table \ref{tab:magnetic_states}).
Using the constraint described in Appendix \ref{magn_constr_section}, it was not possible to stabilize the LS state when the cell volume exceeded approximately 210 \AA$^{3}$. In fact, we can see how the HS state is favored by this expansion, but it also manages to survive under cell compression, although being higher in energy than the LS state.

\subsection{Fe $3d$ orbital occupation of the magnetic ground state} \label{occupation_appendix}

\begin{figure}[!htb]
\centering
\includegraphics[width=0.49\textwidth]{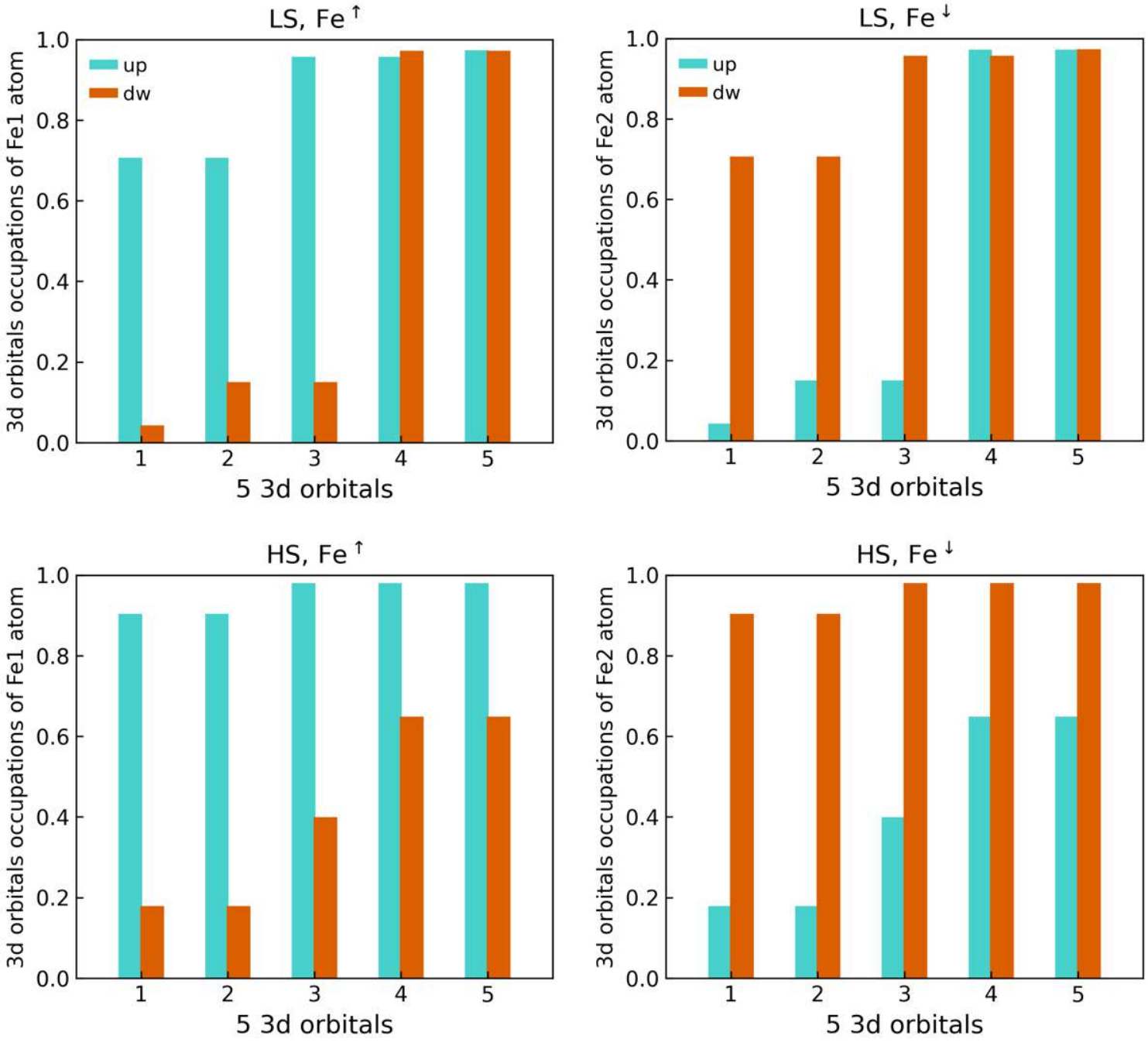}
\caption{Eigenvalues of the occupation matrix for the Hubbard manifold, showing the electronic populations of the five Fe 3$d$ orbitals in the magnetic LS ground state (upper panel) and the related HS state (lower panel). The spin up and spin down channels are shown in cyan and orange, respectively.}
\label{fig:occupationsLS_HS}
\end{figure}

The eigenvalues of the occupation matrix of the Hubbard manifold describing the electronic populations of Fe $3d$ orbitals for the magnetic LS ground state and the related HS state are reported in upper and lower panels of Fig. \ref{fig:occupationsLS_HS}, respectively.
While it is true that Fe and Sb atoms hybridize, we also clearly observe a gap in the eigenvalues of the Fe $d$-orbital occupation matrix between the $e_{g}$ and $t_{2g}$ states in the LS configuration.
\begin{figure*}[!htb]
\centering
\includegraphics[width=\textwidth]{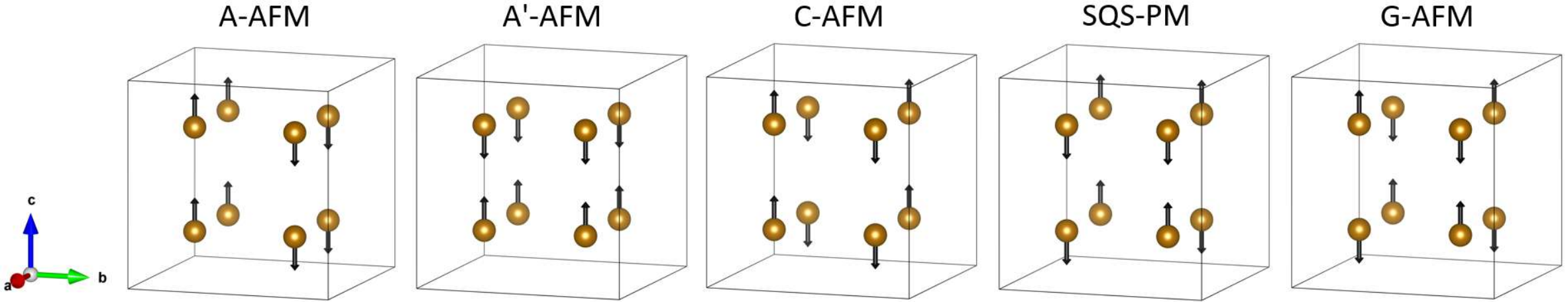}
\caption{From left to right: magnetic moment configurations in the 32-atom conventional cubic (\textit{P}23) cell for the A-, A\textquotesingle-, C-, SQS-, and G-AFM states. Black arrows indicate the direction of the magnetic moments of Fe$^{3+}$ atoms. Gold spheres represent Fe atoms, while Sb atoms are omitted for clarity. All configurations exhibit metallic behavior, except for the SQS-PM state (see Fig. \ref{fig:AFM_configs_bands}).}
\label{fig:magnetic_configs_32_atoms_cell}
\end{figure*}
\begin{figure*}[!htb]
\centering
\includegraphics[width=\textwidth]{./bands_and_dos_compressed.pdf}
\caption{Electronic band structure and density of states of the A- (a), A\textquotesingle- (b), C- (c), G- (d), and SQS-PM (e-f) configurations depicted in Fig. \ref{fig:magnetic_configs_32_atoms_cell}. The spin up (spin down) contribution to the band structure is given by full (dashed) lines. The zero of the energy axis has been set equal to the Fermi level and the highest occupied level for metallic and semiconducting states respectively.}
\label{fig:AFM_configs_bands}
\end{figure*}
Due to this hybridization, the Fe $e_{g}$ orbitals—expected to be unoccupied in a pure crystal field picture—are slightly populated, leading to an increased electron count on the Fe atoms. In fact, the magnetic moment obtained from the difference between the spin-up and spin-down traces of the occupation matrix is 2.01$\mu$\textsubscript{B}, fully consistent with the value of 1.94$\mu$\textsubscript{B} obtained from integration within the muffin-tin sphere as implemented in \texttt{Quantum ESPRESSO}. If the crystal field model were strictly applicable, we would expect the Fe eigenvalues to sum to 5, corresponding to four Fe atoms in a +3 oxidation state (in a LS configuration, each with one unpaired electron), and the Sb eigenvalues to sum to 4, reflecting twelve Sb atoms in a -1 oxidation state, ensuring overall charge neutrality. In the present case, the Fe occupation matrices show that five electrons fully occupy the $t_{2g}$ orbitals, while the $e_{g}$ orbitals, although slightly populated, have significantly lower eigenvalues and are clearly separated by a gap, as noted above. It is important to highlight that the interpretation of occupation matrix eigenvalues in DFT+Hubbard calculations follows the methodology described by \textit{Sit et al.} \cite{sit2011simple}. The slight population of the Fe $e_{g}$ orbitals originates from electron transfer from the Sb atoms, which are not exactly in the -1 oxidation state but rather in an intermediate state between -1 and neutral—corresponding to a $p$-orbital electron count between 3 and 4. For completeness, we also report the HS solution, which lies 1.76 eV higher in energy with respect to the LS ground state and exhibits a significantly larger moment of 2.70$\mu$\textsubscript{B}, che come si vede dai pannelli inferiori di Fig. \ref{fig:occupationsLS_HS}.
\begin{figure*}[!htb]
\centering
\includegraphics[width=\textwidth]{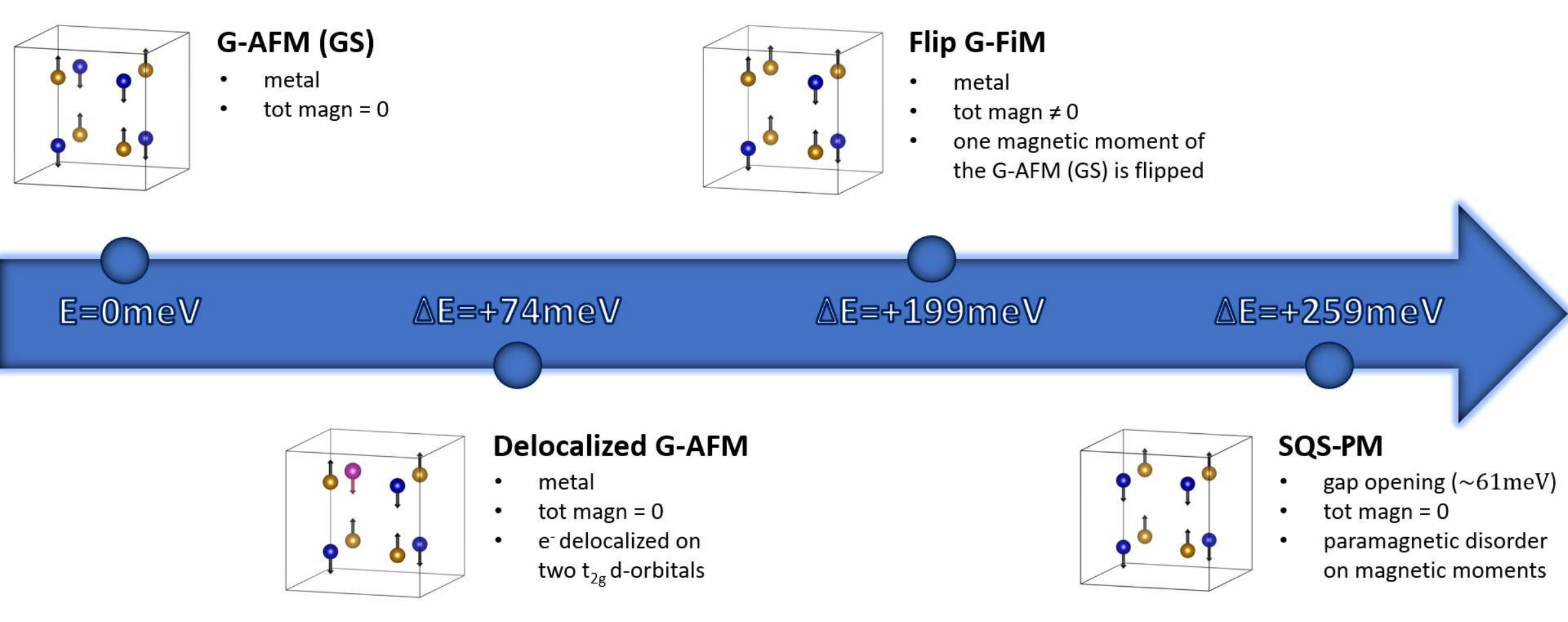}
\caption{Intermediate states between the G-AFM ground state and the SQS-PM state, along with their corresponding energies, magnetic arrangements and total magnetizations. Fe$^{3+}$ atoms with magnetic moments oriented upward and downward are colored gold and blue, respectively, for ease of reference. To differentiate from the others, the Fe$^{3+}$ atom whose atomic orbitals exhibit the internal delocalization of a LS electron is colored in purple.}
\label{fig:very_first_excitations}
\end{figure*}
This enhanced moment arises from the absence of a clear gap in the spin-up orbital occupations. 

In summary, although Fe–Sb hybridization is clearly present, our interpretation based on the explicit eigenvalues of the Fe 
$d$ and Sb $p$ orbital occupation matrices remains valid for distinguishing between the lower-energy LS states and the HS ones.

\subsection{Magnetic configurations} \label{magnetic_configs_appendix}

In this section we show the detailed arrangement of local magnetic moments directions in the 32-atom cubic (\textit{P}23) cell. In Fig. \ref{fig:magnetic_configs_32_atoms_cell} we provide the magnetic arrangement of the different standard AFM configurations; the related electronic band structures computed using Hubbard obtained self-consistently for the AFM C3 state are also shown in Fig. \ref{fig:AFM_configs_bands}. Here the lattice parameter of the cubic cell is the one of the G-AFM optimized geometry equal to 9.206$\rm{\AA^{3}}$. Interestingly, the C-AFM state is higher in energy than the SQS-PM one but remains very close, at just 0.01 eV/Fe (see Table \ref{tab:magnetic_states_32_atoms}). The C-AFM state also exhibits clear non-degeneracies in the spin up and spin down bands along the same segment of the Brillouin zone where the SQS-PM state shows them.
\begin{figure}[!htb]
\centering
\includegraphics[width=0.49\textwidth]{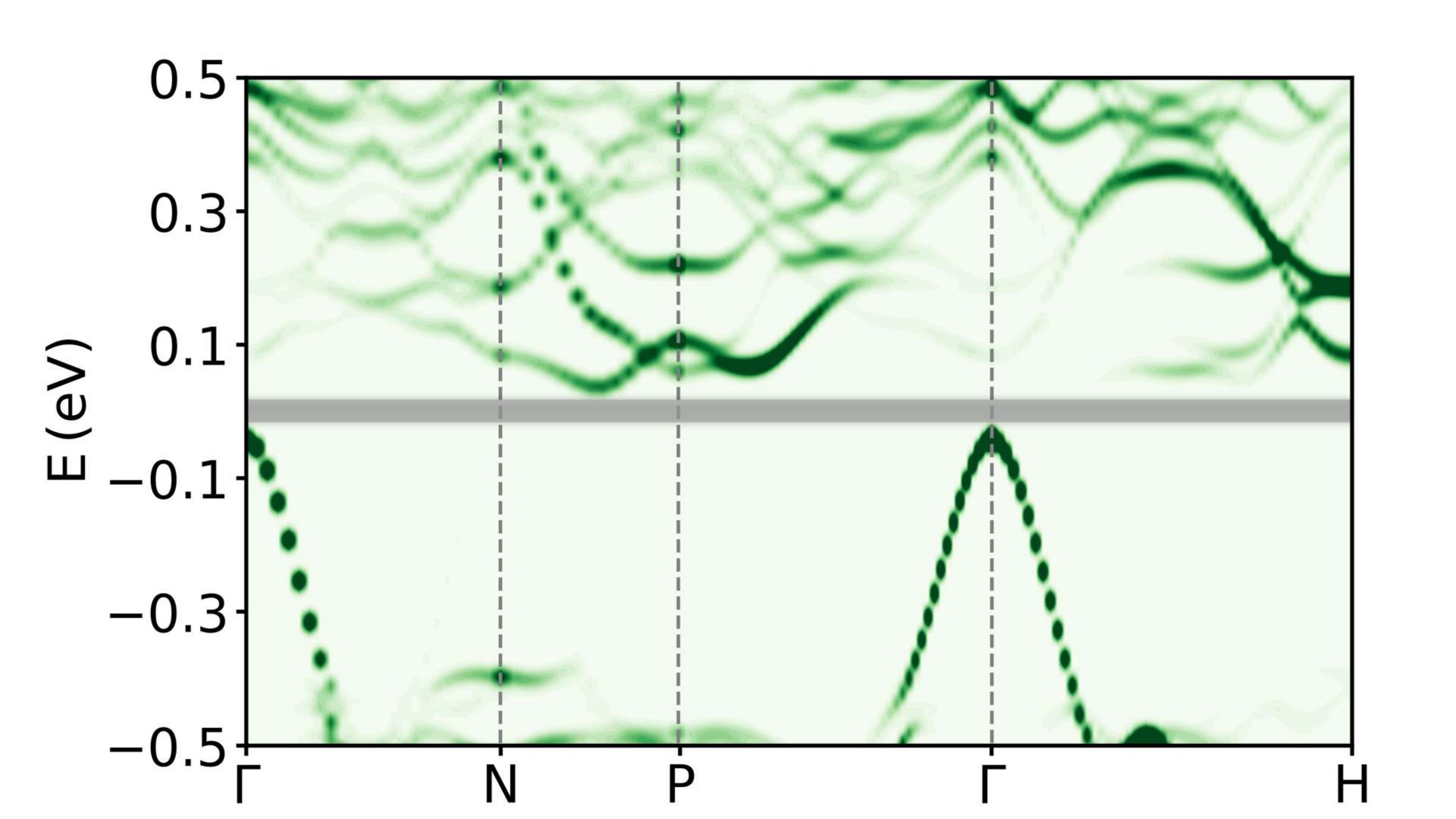}
\caption{Electronic band structure of the SQS-128 (C3) state (see Table \ref{tab:magnetic_states}). The band structure is calculated along the main symmetry directions. The supercell band structure of the SQS-128 (C3) state is unfolded into the primitive unit cell by using the python version of the BandUP code \cite{bandup1,bandup2,bandup3}. The color scale represents the weight assigned to each point of the mapping between the primitive cell and supercell as given in Ref. \cite{bandup1}. Here the grey shaded area represents the band gap of $\sim38$meV.}
\label{fig:sqs-128}
\end{figure}
This state is, in fact, the one that follows the SQS-PM state, whose configuration is identified by the \texttt{ATAT} simulation as the one exhibiting the second highest degree of disorder in the orientation of magnetic moments for the given 32-atom cell. Fig. \ref{fig:very_first_excitations} summarizes the states found in the energy range between the G-AFM ground state and the SQS-PM state, along with their corresponding energies and total magnetizations. All intermediate states between the G-AFM ground state and the SQS-PM state exhibit metallic behavior. The electronic band structure of the flip G-FiM state (not reported) shows non-degenerate spin up and spin down bands along all high-symmetry paths in the Brillouin zone, as expected for ferrimagnetic systems.

\section{SQS state in larger supercells} \label{128_atom_section}

The electronic band structure of the SQS state in the 128-atom supercell is shown in Fig. \ref{fig:sqs-128}.
\begin{figure}[!htb]
\centering
\includegraphics[width=0.49\textwidth]{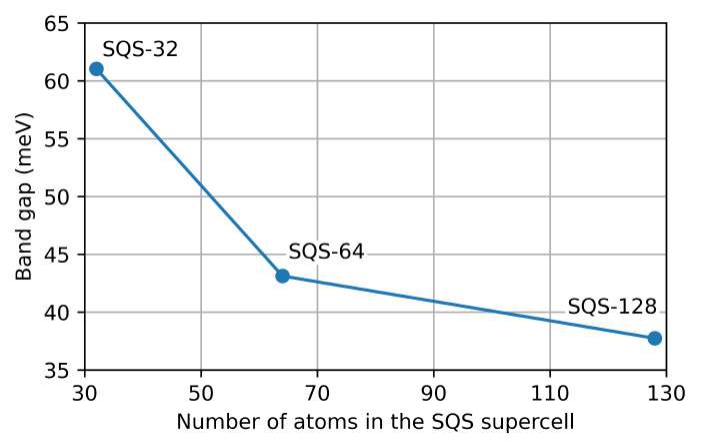}
\caption{Evolution of the calculated band gap as a function of the SQS supercell size. Results are shown for SQS supercells containing 32, 64, and 128 atoms, highlighting the gradual convergence of the band gap with increasing configurational sampling.}
\label{fig:SQS_gap_convergence}
\end{figure}
To assess the convergence of the electronic band gap with respect to configurational sampling, we also examine an SQS configuration constructed in a $1\times1\times2$ 64-atom supercell of the 32-atom conventional cubic cell. As shown in Fig. \ref{fig:SQS_gap_convergence}, the resulting band gaps are approximately 61 meV (SQS-32), 43 meV (SQS-64), and 38 meV (SQS-128), revealing a systematic reduction of the gap with increasing supercell size. This trend suggests that the SQS-128 result shown in Fig. \ref{fig:sqs-128} is close to convergence. Importantly, achieving band gaps on the order of a few tens of meV already represents a substantial improvement over typical local and semi-local DFT (LDA/GGA) predictions, whose errors are commonly on the order of electron-volts. This improvement is enabled by the fully ab-initio, self-consistent DFT+$U+V$ approach which, although not specifically designed to deliver quantitatively accurate band gaps, captures the essential physics of magnetic fluctuations responsible for opening the gap.

\section{Computational details}

In this work we use Hubbard-corrected density-functional theory \cite{anisimov1991band,dudarev1998electron} to overcome the limitations of the local/semilocal DFT (e.g. LDA and GGA) \cite{cohen2008insights} in dealing with the strongly-localized self-interacting 3$d$ electrons in iron \cite{kulik2006density,timrov2022hp}. In particular, we include both the on-site $U$ and the intersite $V$ terms: these latter link the Fe atoms with their surrounding ligands (six Sb atoms coordinated in octahedral complexes with one Fe at the center). Moreover, we use Löwdin orthogonalized atomic wave functions as projectors, as described in Ref. \cite{timrov2021self}, and we compute Hubbard $U$ and $V$ parameters from linear-response density-functional perturbation theory (DFPT) \cite{cococcioni2005linear}, which makes the approach fully parameter-free and avoids any empirical treatment.\\
We performed DFT+U+V calculations using the \texttt{Quantum ESPRESSO} distribution \cite{quantum_espresso} with projector-augmented-wave \cite{PAW} and ultrasoft pseudopotentials for Fe and Sb, respectively, as prescribed by standard solid-state pseudopotentials (SSSP efficiency, version 1.3) library \cite{SSSP,SSSP2,SSSP3}. The Perdew-Burke-Emzerhof approximation for solids (PBEsol) is employed for the exchange-correlation functional \cite{perdew2008restoring}. Convergence of total energy, forces, and pressure with respect to $k$-mesh and kinetic energy cutoff is achieved using a $6\times6\times6$ $k$-points grid, and 60\,Ry plane wave cutoff respectively. The phonon calculations are performed within a finite difference approach by using the \texttt{Phonopy}
package \cite{phonopy}; 3$\times$3$\times$3 supercell with a 2$\times$2$\times$2 $k$-mesh is used to compute forces on displaced configurations. The SQSs used for the distributions of the atomic magnetic moments are generated using the \texttt{mcsqs} code from the alloy theoretic automated toolkit (\texttt{ATAT}) \cite{van2002alloy}.\\
For the Wannierization, we first performed a non-self-consistent (NSCF) calculation including 800 bands. In constructing the Wannier functions, we excluded 152 semicore states and retained a total of 168 Wannier functions: nine orbitals ($s$, $p$, $d$) for each of the eight Fe atoms, and four orbitals ($s$, $p$) for each of the 24 Sb atoms. The disentanglement window was set to –2 to 60 eV, and the frozen window to –2 to 14 eV.

%\bibliographystyle{apsrev4-1}
%\bibliography{bib}

%

\end{document}